\documentclass[aps,showpacs,nofootinbib,floatfix]{revtex4}
\usepackage{graphicx}
\usepackage{amsmath, amssymb, bm, graphicx, graphics, color,mathrsfs,ulem}

\newcommand{\be}{\begin{equation}}
\newcommand{\ee}{\end{equation}}
\newcommand{\ben}{\begin{eqnarray}}
\newcommand{\een}{\end{eqnarray}}

\newcommand{\cO}{{\cal O}}

\newcommand{\cE}{{\cal E}}

\newcommand{\cB}{{\cal B}}

\newcommand{\p}{\partial}
\newcommand{\na}{\nabla}

\newcommand{\tG}{\tilde G}

\newcommand{\tQ}{\tilde Q}

\newcommand{\ep}{\epsilon}

\newcommand{\tB}{{\tilde B}}

\newcommand{\tkappa}{{\tilde \kappa}}

\newcommand{\tmu}{{\tilde \mu}}

\begin{document}
\title{Two interacting current model of holographic Dirac fluid in graphene}


\author{Marek Rogatko} 
\email{rogat@kft.umcs.lublin.pl,
marek.rogatko@poczta.umcs.lublin.pl }
\author{Karol I. Wysokinski}
\email{karol.wysokinski@umcs.pl}
\affiliation{Institute of Physics \protect \\
Maria Curie-Sklodowska University \protect \\
20-031 Lublin, pl.~Marii Curie-Sklodowskiej 1, Poland}

\date{\today}

\begin{abstract}
The electrons in graphene for energies close to the Dirac point 
have been found to form strongly interacting fluid. Taking this fact into account we
have extended previous work on the transport properties of graphene
by taking into account possible interactions between the currents and 
 adding the external magnetic field directed perpendicularly to the graphene sheet.
The perpendicular magnetic field $B$ severely modifies the transport parameters. In the
present approach the quantization of the spectrum and formation of Landau levels is ignored.
Gauge/gravity duality has been used in the probe limit. The dependence on the charge density 
of the Seebeck coefficient and thermo-electric parameters $\alpha^{ij}$ nicely agree with
recent experimental data for graphene. 
The holographic model allows for the interpretation of one of the  fields representing the currents
as resulting from the 
{\it dark matter sector}. For the studied geometry with electric field
perpendicular to the thermal gradient the effect of {\it dark sector} has been found to modify
the transport parameters but mostly in a quantitative way only. This makes difficult the detection of this
elusive component of the Universe by studying transport properties of graphene.
\end{abstract}



\maketitle


\section{Introduction}
The crossroads between gravity theory and condensed matter physics have recently become 
an intense field of research with at least two-fold goal. On one side, the expectation
of the condensed matter community is that the approach providing strong coupling analysis
 of problems will shed some light on those aspects being difficult
to access by other means \cite{erdbook,zaanen-book}. On the other hand, such studies 
can shed the light on the question whether the holographic approach is able
to describe real phenomena observed in experiments.

The exploit of the gauge/gravity correspondence \cite{mal,wit98,gub98} in studying strongly correlated systems 
resulted, among others, in establishing the lower bound $\hbar/4\pi$ on the ratio of the shear 
viscosity $\eta_s$ to entropy density $s$ in holographic fluid \cite{kovtun2005}. This interesting
result  has contributed  to the  deeper understanding of the state of strongly interacting 
quark-gluon plasma obtained at RHIC \cite{rom07}-\cite{mat07}. 
Related studies based on the gauge/gravity duality \cite{har07,luc16} have also triggered the shear viscosity 
measurements in the  ultra-cold Fermi gases \cite{cao11}, and more recently in the condensed matter systems such as  
graphene~\cite{crossno2016,bandurin2016} and strongly correlated  oxide \cite{moll2016}. 
The comprehensive discussion of this novel set of experiments is given in~\cite{zaanen2016}.

{Recently, a great resurgence of the interests in holographic lattice 
studies of the thermoelectric DC transport has been observed.
Breaking of the translation invariance provides the mechanism 
of momentum dissipation in the underlying field theory and disposes to the finite values of 
holographic DC
kinetic coefficients including thermoelectric matrix elements.}

{The number of results have already been obtained by this technique for a similar model of dissipation 
and valid in principle for arbitrary value of temperature and the strength of momentum dissipation. 
Namely, the massive gravity electrical conductivity was analyzed in \cite{bla13}-\cite{dav13} 
and the consecutive  generalization to the lattice models appeared \cite{bla14}-\cite{don14a}. 
The linear axions disturbing  the translation invariance were elaborated 
in \cite{and14}, while the thermal conductivities were calculated in \cite{don14b}-\cite{amo15}.}

{On the other hand, it was shown that for Einstein-Maxwell scalar field gravity, 
the thermoelectric DC conductivity of the dual field theory can be achieved 
by considering a linearized Navier-Stokes equations on the black hole event horizon \cite{don15}-\cite{don16}. 
The studies in question were generalized to higher derivative gravity, which emerged 
due to the perturbative effective expansion of the string action \cite{don17}. 
The exact solution for Gauss-Bonnet-Maxwell scalar field theory for holographic 
DC thermoelectric conductivities with momentum relaxation was given in \cite{che15}}.

{The  important ingredient in studying transport properties is a magnetic field, 
which is essential in such phenomena like quantum Hall, the Nerst and other effects.  
The researches in this direction were conducted in \cite{bla15}-\cite{kim15}. 
Recently, the very important holographic generalization of the hydrodynamic approach \cite{foster2009} 
appeared \cite{seo17}, where the holographic model of strongly coupled plasma with two distinct conserved $U(1)$-gauge currents was presented, in order to
describe the nature of graphene. The very good agreement with the existing experimental data was achieved.}

In our paper we shall study {some generalization of the aforementioned model \cite{seo17}. Namely, we shall elaborate
the transport properties of 2+1 dimensional strongly coupled quantum fluid
in a graphene under the influence of weak ($i.e.,$ non-quantizing) perpendicular magnetic field and in the presence 
of the second $U(1)$-gauge field. Our model assumes the interaction between
both  fields responsible for the adequate  currents. 
The main objective of our work is to find the influence of $\alpha$-coupling constant of the fields in question on the transport properties of the holographic model of graphene.}

It has to be recalled that the geometry of the system is crucial and has to be carefully
analyzed when comparing the results with experimental data on graphene.

The paper is organized as follows. In Sec.II we present the holographic model and 
discuss the adequate perturbations needed to find the currents
in the system. One also pays attention to the generalization of the Sachdev 
model of holographic Dirac fluid with two interacting currents.
In Sec.III we find the transport coefficients for the underlying holographic 
model with the influence of magnetic field. Sec.IV and V tackle the four-dimensional 
dyonic black hole with two $U(1)$-gauge fields and the transport and kinetic coefficients for the spacetime 
of black hole in question. In Sec.VI we discuss our results in the light of 
the recent experiments on graphene and elaborate
the dependence of $\alpha$-coupling constant on Hall angle. Sec.VII is devoted to the conclusions, as well as, 
we  provide there, the discussion of the other possible interpretation of the model, as a model
of {\it dark matter} sector.

\section{Holographic model}

In this section we shall tackle the problem of the holographic set-up. 
{It has been argued that the hydrodynamical models as suggested in \cite{har07, luc16} lead to a better agreements with the observations 
but still there exist a room for improvements.
In \cite{seo17} the holographic model of the two conserved $U(1)$-gauge currents with momentum dissipation envisaging the weak point-like disorder, was introduced to describe Dirac fluid. The main idea
standing behind the introducing a new current was that it could enhance the transport of the heat relative to its charge.}

{In the present paper we propose some generalization of the aforementioned model, 
considering two interacting $U(1)$-gauge fields. The main objective in our research will be to find the influence of the
field coupling constant on the transport properties of the system in question.}

The gravitational background for the 
holographic model in $(3+1)$-dimensions with the two interacting $U(1)$-gauge fields is taken in the form 
\be
S = \int \sqrt{-g} d^4 x  \big( R + \frac{6}{L^2} - \frac{1}{2} \na_\mu \phi_i \na^\mu \phi^i
- \frac{1}{4} F_{\mu \nu} F^{\mu \nu} 
- \frac{1}{4}B_{\mu \nu} B^{\mu \nu}
 - \frac{\alpha}{4} F_{\mu \nu} B^{\mu \nu} \big),
\label{sgrav} 
\ee
where $F_{\mu \nu} = 2 \nabla_{[ \mu} A_{\nu ]}$ stands for the ordinary Maxwell field strength tensor, while
the second $U(1)$-gauge field $B_{\mu \nu}$ is given by $B_{\mu \nu} = 2 \nabla_{[ \mu} B_{\nu ]}$. $\alpha$ is a coupling constant between two gauge fields.

The justifications of such kind of models can be acquitted from the top-down perspective \cite{ach16}, starting from
the string/M-theory. This fact is important in the holographic attitude, since the theory 
in question is a fully consistent quantum theory and 
 guarantees  that any phenomenon described by the top-down theory is physical.
In the action (\ref{sgrav}) the second gauge field is bounded with some hidden sector \cite{ach16}.
The term which depicts interaction of visible (Maxwell field) sector and the hidden $U(1)$-gauge field is called the {\it kinetic mixing term}.
For the first time it was used in \cite{hol86}, in order to describe the existence and subsequent integrating out of heavy bi-fundamental fields charged under the $U(1)$-gauge groups.
In general, such kind of terms arise in the theories that have in addition to some visible gauge group an
additional one, in the hidden sector. The compactified string or M-theory solutions generically possess
hidden sectors (containing at a minimum, the gauge
fields and gauginos, due to the various group factors included in the gauge group symmetry of the hidden sector).
The hidden sector contains states in the low-energy effective theory which are uncharged under the the Standard 
Model gauge symmetry groups. They are charged under their own groups. Hidden sectors interact with the visible ones via gravitational interaction. 
In principle one can also think out other portals to our visible sector. This interesting problem was discussed in \cite{portal1, portal2}

One can also notice, that many extensions of the Standard Model also contain hidden sectors that have no renormalizable interactions with particle of the model in question. 
The realistic embeddings of the Standard Model in $E8 \times E8$ string theory, as well as, in type I, IIA, or IIB open string theory with branes, require the existence of the hidden sectors for the consistency and supersymmetry breaking \cite{abe08}. 
The most generic portal emerging from the string theory is the aforementioned {\it kinetic mixing} one.

The {\it kinetic mixing term} can contribute significantly and dominantly to the supersymmetry breaking mediation
\cite{abe04,die97}, ensuing in the contributions to the scalar mass squared terms proportional to their hypercharges. 
The mediation of supersymmetry breaking, in models involving stacks of D-brane and anti D-brane, producing  a kinetic mixing term of $U(N)$-groups, was presented in \cite{abe04}.

Generally, in string phenomenology \cite{abe08} the dimensionless kinetic mixing term parameter {$\alpha$} 
can be produced at an arbitrary high energy scale and it does not deteriorate 
from any kind of mass suppression from the messenger introducing it. This fact is of a great importance from the experimental point of view, due to the fact
that its measurement can provide some interesting features of 
high energy physics beyond the range of the contemporary colliders.

The mixing term of two gauge sectors are typical for states for open 
string theories, where both $U(1)$-gauge groups are advocated by D-branes that are separated in extra dimensions. It happens
in supersymmetric Type I, Type IIA, Type IIB models. It results in the existence of massive open strings 
which stretch between two D-branes in question. It accomplishes the scenario of the connection of 
different gauge sectors. It can be realized by M2-branes wrapped on surfaces which intersect 
two distinct codimension four orbifolds singularities (they correspond (at low energy) to massive 
particles which are charged under both gauge groups).
Some generalizations of this statement to M, F-theory and heterotic string theory are also known.

On the other hand, the model with two coupled vector fields, was also implemented 
in a generalization of p-wave superconductivity, for the holographic model of ferromagnetic 
superconductivity \cite{amo14} and, without coupling $\alpha$, for the description 
of the thermal conductivity in graphene \cite{seo17}.

The equations of motion  obtained from the variation of the action $S$ with respect
to the metric, the scalar and gauge fields imply
\ben 
G_{\mu \nu} - \frac{3 g_{\mu \nu}}{L^2} &=& T_{\mu \nu}(\phi_i) + T_{\mu \nu}(F) + T_{\mu \nu}(B) 
+ \alpha T_{\mu \nu}(F,B),\\ \label{ff1}
\na_{\mu}F^{\mu \nu} &+& \frac{\alpha}{2}~\na_\mu B^{\mu \nu} = 0,\\
\na_{\mu}B^{\mu \nu} &+& \frac{\alpha}{2}~\na_\mu F^{\mu \nu} = 0,\\
\na_\mu \na^\mu \phi_i &=& 0,
\een
where the energy momentum tensors for the adequate  fields are provided by
\ben
T_{\mu \nu} (\phi_i) &=& \frac{1}{2} \na_{\mu} \phi_i \na_\nu \phi_i - \frac{1}{4}~g_{\mu \nu}~\na_\delta \phi_i \na^\delta \phi_i ,\\
T_{\mu \nu}(F) &=& \frac{1}{2}~F_{\mu \delta}F_{\nu}{}^{\delta} - \frac{1}{8}~g_{\mu \nu}~F_{\alpha \beta}F^{\alpha \beta},\\
T_{\mu \nu}(B) &=& \frac{1}{2}~B_{\mu \delta}B_{\nu}{}^{\delta} - \frac{1}{8}~g_{\mu \nu}~B_{\alpha \beta}B^{\alpha \beta},\\
T_{\mu \nu}(F,~B) &=& \frac{1}{2}~F_{\mu \delta}B_{\nu}{}^{\delta} - \frac{1}{8}~g_{\mu \nu}~F_{\alpha \beta}B^{\alpha \beta}.
\een
One supposes that the scalar fields depend on the three spatial coordinates, i.e., 
$\phi_i (x_\alpha) = \beta_{i \mu} x^\mu = a_i x + b_i y$.
The dependence will be of the same form for all the coordinates,  which means that $a_i =b_i =\beta$.

In the considered holographic model, we propose the ansatze for the gauge fields given by
\ben
A_\mu(r) ~dx^\mu &=& a(r) ~dt + \frac{B}{2} ~(xdy-ydx),\\
B_\mu(r)~dx^\mu  &=& b(r)~dt,
\een
where by $B$ is a background magnetic field. 

The general spacetime which will be consistent with the above choice implies
\be
ds^2 = - f(r)dt^2 + \frac{dr^2}{f(r)} + r^2 (dx^2 + dy^2 ).
\label{sp}
\ee
In order to find the thermoelectric and DC-conductivities one should find the radially independent quantities in the bulk that can be identified with the adequate boundary currents \cite{don14,don14a,don14b,kim15}.

First let us suppose that $k_\alpha = (\p/\p t)_\alpha$ is a timelike Killing vector field. Because of the fact that we are considering the static spacetime the spacelike hypersurfaces
are orthogonal to the orbits of the isometry generated by the Killing vector field in question. The general properties of the Killing vector field and gauge fields in visible and hidden
sectors, enable us to define the two-form which implies
\ben \nonumber
\tG^{ \nu \rho} &=& \na^\nu k^\rho + \frac{1}{2} \Big( k^{[\nu}F^{\rho] \alpha}A_\alpha \Big) + \frac{1}{4} \Big[ \Big( \psi - 2 \theta_{(F)} \Big)~F^{\nu \rho} \Big] \\ \nonumber
&+& \frac{1}{2} \Big( k^{[\nu}B^{\rho] \alpha}B_\alpha \Big) + \frac{1}{4} \Big[ \Big( \chi - 2 \theta_{(B)} \Big)~B^{\nu \rho} \Big] \\ 
&+& \frac{\alpha}{4} \Big[ \Big( k^{[\nu}B^{\rho] \alpha}A_\alpha \Big) + \Big( k^{[\nu}F^{\rho] \alpha}B_\alpha \Big) \Big] \\ \nonumber
&+& \frac{\alpha}{8}\Big[ \Big( \psi - 2 \theta_{(F)} \Big)~B^{\nu \rho} \Big] + \frac{\alpha}{8} \Big[ \Big( \chi - 2 \theta_{(B)} \Big)~F^{\nu \rho} \Big].
\een
where we have set   $\psi,~\chi,~\theta_{(F)},~\theta_{(B)}$  the following relations:
\ben
\psi &=& E_\alpha x^\alpha, \qquad \theta_{(F)} = - E_\alpha x^\alpha - a(r),\\
\chi &=& B_\beta x^\beta, \qquad \theta_{(B)} = - B_\beta x^\beta - b(r),
\een
where $\alpha,~\beta = x,~y$.  In the above equations $E_a$ is the Maxwell electric field while $B_a$ is 'electric' field is bounded with the hidden sector gauge field. 
As it can be deduced from the definition, $\tG_{\alpha \beta}$ tensor is antisymmetric and fulfills the following:
\be
\p_\rho \big( 2~\sqrt{-g}~\tG^{\nu \rho} \big) = - 2 \frac{\Lambda~\sqrt{-g}~k^\nu}{d-2},
\label{2ff}
\ee
where $d$ stands for the dimensionality of the spacetime, while $\Lambda$ is the cosmological constant.\\
A close inspection of (\ref{2ff}) reveals that the right-hand side 
is equal to zero if one considers the Killing vector $k^\nu$
with the index different from the connected with time coordinate. 
In our considerations we shall use the two-form given by $2 \tG_{ \nu \rho} $, i.e.,
the heat current will be defined as $Q^i = 2~\sqrt{-g} \tG^{ ri} $.

On the other hand, having in mind equations of motion for gauge fields, one finds the adequate conserved charges in the $r$-direction
\ben
\tQ_{(F)} &=& \sqrt{-g}~\big(F^{rt} + \frac{\alpha}{2} B^{rt} \big) = Q_{(F)} + \frac{\alpha}{2} Q_{(B)},\\
\tQ_{(B)} &=& \sqrt{-g}~\big(B^{rt} + \frac{\alpha}{2} F^{rt} \big) = Q_{(B)} + \frac{\alpha}{2} Q_{(F)},
\een
where we set $
Q_{(F)} =  r^2~a'(r),~Q_{(B)}=  r^2~b'(r).$

In order to find the conductivities for the background in question, one takes into account small perturbations around the background solution obtained from Einstein equations of motion.
The perturbations imply
\ben \label{f1}
\delta A_i &=& t~\Big( - E_i + \xi_i~a(r) \Big) + \delta a_i(r),\\
\delta B_i &=& t~\Big( - B_i + \xi_i~b(r) \Big) + \delta b_i(r),\\
\delta G_{ti} &=& t~\Big( - \xi_i~f(r) \Big) + \delta g_{ti}(r),\\
\delta G_{r i} &=& r^2~\delta g_{ri}(r),\\ \label{f5}
\delta \phi_i  &=& \delta \phi_i(r),
\een
where $t$ is time coordinate. We put $i= x,~y$, and denote the temperature gradient 
by $\xi_i = - \na_i T/T$.

However, the presence of magnetization causes that one should take into account the non-trivial fluxes connected with the non-zero components $B$. The linearized
equations describing can be written in the form as
\ben \nonumber
0 &=& \p_{M} \big[ \sqrt{-g} \big( F^{iM} + \frac{\alpha}{2} B^{iM} \Big) \big] = \p_r  \big[ \sqrt{-g} \big( F^{i r} + \frac{\alpha}{2} B^{i r} \big) \big] \\ 
&+& \p_t  \big[ \sqrt{-g}~\big( F^{i t} + \frac{\alpha}{2} B^{i t} \big) \big] ,
\een
and for the other gauge field equation of motion
\ben \nonumber
0 &=& \p_{M} \big[ \sqrt{-g}~\big( B^{iM} + \frac{\alpha}{2} F^{iM} \big) \big] = \p_r  \big[ \sqrt{-g}~\big( B^{i r} + \frac{\alpha}{2} F^{i r} \big) \big] \\
&+& \p_t  \big[ \sqrt{-g}~\big( B^{i t} + \frac{\alpha}{2} F^{i t} \big) \big] .
\een

Because of the fact that {\it electric currents} are r-independent, we shall evaluate them on the black object event horizon. Integrating
the above relations we arrive at the currents at the boundary of $AdS_4$
\ben
J^i_{(F)}(\infty) &=& J^i_{(F)}(r_h) + \frac{B}{2} \ep^{ij}~\xi_j~ \Sigma_{(1)},\\
J^i_{(B)}(\infty ) &=& J^i_{(B)}(r_h) +  \frac{\alpha}{2} \frac{B}{2}~\xi^i~\Sigma_{(1)} ,
\een
where $\Sigma_{(1)} = \int_{r_h}^{\infty} dr'~\frac{1}{r'^2}$ and $\ep^{ij}$ is a two dimensional
anti-symmetric tensor, $\ep^{ij} = -\ep^{ji}$. The symbol $\ep^{ij}$ is uniquely determined by its symmetry properties up to a constant, we choose that $\ep_{yx} = - \ep_{xy} =1$.

The heat current at the linearized order implies
\be
Q^i(r) = 2~\sqrt{-g}\na^r k^i - a(r)~J^i_{(F)}(r) - b(r)~J^i_{(B)}(r),
\ee
The heat current is subject to the relation $\p_\mu [2\sqrt{-g} \tG^{\mu \nu}] = 0$, in the absence of a thermal gradient. But the existence of magnetization currents enforced that we have the following 
equations:
\ben \nonumber
\p_r [ 2\sqrt{-g}\tG^{rx}] &=& -\p_t [2\sqrt{-g}\tG^{tx}] - \p_y [ 2\sqrt{-g}\tG^{yx} ] \\
&-& a(r)  J^x_{(F)}(\infty)  - b(r) J^x_{(B)}(\infty) ,\\ \nonumber
\p_r [2 \sqrt{-g}\tG^{ry}] &=& -\p_t [2\sqrt{-g}\tG^{ty}] - \p_y [ 2\sqrt{-g}\tG^{xy} ] \\
&-& a(r)  J^y_{(F)}(\infty)  - b(r) J^y_{(B)}(\infty).
\een
In order to achieve the radially independent form of the current, one ought to add additional terms to get rid of the aforementioned fluxes.
The considered quantity should obey $\p_i \tQ^i = 0$,  then one has to have
\ben \nonumber
\tQ^i (\infty) &=& Q^i(r_h) + \frac{B}{2}~\ep^{ij}~E_j \Sigma_{(1)} - B~\ep^{ij}\xi_j \Sigma_{(a)} \\ 
&-& \frac{\alpha}{2} B~\ep^{ij}~B_j \Sigma_{(b)} 
+ \frac{\alpha}{4} B~\ep^{ij}~B_j \Sigma_{(1)},
\een
where we have denoted $
\Sigma_{(a)} = \int_{r_h}^{\infty} dr'~\frac{a(r')}{r'^2}, ~\Sigma_{(b)} = \int_{r_h}^{\infty} dr'~\frac{b(r')}{r'^2}.$
We have obtained three boundary currents $J^i_{(F)}(\infty),~J^i_{(B)}(\infty)$ and $\tQ^i(\infty)$, which can be simplified by imposing the regularity conditions at the black brane horizon.
Namely, they imply the following:
\ben
\delta a_i(r) &\sim& - \frac{E_i}{4~\pi~T} \ln (r-r_h) + \dots,\\
\delta b_i(r) &\sim& - \frac{B_i}{4~\pi~T} \ln (r-r_h) + \dots,\\
\delta g_{ri}(r) &\sim& \frac{1}{r_h^2}~\frac{\delta g^{(h)}_{ti}}{f(r_h)} + \dots,\\
\delta g_{ti}(r) &\sim& \delta g^{(h)}_{ti}+ \cO(r-r_h) + \dots,\\
\delta \phi_i(r) &\sim & \phi_i(r_h) +  \cO(r-r_h) + \dots,
\een
where $T=1/4 \pi~\p_r f(r)\mid_{r=r_h}$ is the Hawking temperature of the black brane in question.


\subsection{Generalization of the Sachdev model of the Dirac fluid}

In this subsection we assume that one has no magnetic field in order to confront predictions of our model with the one described in \cite{seo17}. To begin with,
let us define thermoelectric forces for the visible and hidden sector fields as
\ben
\cE_i = E - \na_i \Big(\frac{\mu_F}{T} \Big),\\
\cB_j = \tB - \na_j \Big( \frac{\mu_B}{T} \Big).
\een
The total electric current constitutes the of the currents for the visible sector gauge field $J_{(F)}$ and for the hidden sector one $J_{(B)}$
\be
J =  J_{(F)} + J_{(B)} =  \sigma_{F j} \cE^j + \sigma_{F a} \cB^a +  \sigma_{B j} \cE^j + \sigma_{B a} \cB^a .
\ee
On the other hand, electric conductivity is given by the relation
\be
\sigma = \frac{\p J}{\p E} + \frac{\p J}{\p \tB} = \sigma_{FF} + \sigma_{FB} + \sigma_{BF} + \sigma_{BB}.
\ee
Let us restrict our considerations to x-direction, then
one receives the boundary currents in terms of the external sources like $E,~\tB,~\tQ_{(F)},~\tQ_{(B)}$, provided by
\ben
J_{(F)} (\infty) &=& E~\Big(1 + \frac{\tQ_{(F)}^2}{\beta^2} \Big) + \tB~\Big( \frac{\alpha}{2} + \frac{\tQ_{(F)} \tQ_{(B)}}{\beta^2} \Big) + \frac{4 \pi T~r_h^2}{\beta^2} \tQ_{(F)}~\xi,\\ \label{jf}
J_{(B)}(\infty) &=&  \tB~\Big(1 + \frac{\tQ_{(B)}^2}{\beta^2} \Big) + E~\Big( \frac{\alpha}{2} + \frac{\tQ_{(B)} \tQ_{(F)}}{\beta^2} \Big) + \frac{4 \pi T~r_h^2}{\beta^2} \tQ_{(B)}~\xi,\\ \label{qq}
\tQ (\infty) &=& \frac{4 \pi T~r_h^2}{\beta^2} \tQ_{(F)}~E + \frac{4 \pi T~r_h^2}{\beta^2} \tQ_{(B)}~\tB + \frac{16 \pi^2 T^2 r_h^4}{\beta^2}~\xi.
\een
The above relations can be rewritten in a more compact form. Namely, in the matrix form they are given by
\be
\left ( \begin{array}{ccc}
\sigma_{FF} & ~~~\sigma_{FB} &~~~\alpha_{F} T \\
\sigma_{BF}&~~~ \sigma_{BB}&~~~ \alpha_B T \\
\alpha_F T &~~~ \alpha_B T &~~~ \tkappa T
\end{array} \right)
\left( \begin{array}{c}
E \\ \tB \\ \xi \end{array} \right) = \left( \begin{array}{c} J_{(F)}\\ J_{(B)}\\ \tQ \end{array} \right).
\ee
From the equation (\ref{jf})-(\ref{qq}) it can be easily seen that the transport coefficients are real, symmetric and the Onsager relations are fulfilled.

Assuming that the $U(1)$-gauge charges are bounded by the relation
\be
Q_{(B)} = g~Q_{(F)},
\label{bf}
\ee
we arrive at the following equation for the electric conductivity
\be
\sigma = \sigma_0~\Big[  1+  \frac{1}{2 \beta^2} (1+g)^2 (1+ \frac{\alpha}{2}) ~Q_{(F)}^2 \Big],
\label{si}
\ee
where we have denoted $\sigma_0 = 2 + \alpha$. 
Moreover the assumption (\ref{bf}) enables us to write
\be
\tQ_{(F)} = \Big( 1 + \frac{\alpha}{2}g  \Big) Q_{(F)}, \qquad \tQ_{(B)} = \Big( g + \frac{\alpha}{2} \Big) Q_{(F)}.
\ee
If we denote by $Q = Q_{(F)} + Q_{(B)}$, then $Q_{(F)} = Q/(1+g)$. 
Just it leads to the conclusion that in the equation (\ref{si}), we have no dependence on $g$ and
$Q$ has been earlier \cite{seo17} identified with the charge density $n$ in graphene.

Let us find the ratio of the electric conductivity responsible for the two-current 
interaction and electric conductivity without mutual influence. The relation is provided by
\be
\frac{\sigma(\alpha) }{\sigma(0)} = \Big(1 +\frac{\alpha}{2} \Big) \Big[1 + \frac{\alpha~Q^2}{4 \beta^2 \bigg(1 + \frac{Q^2}{2 \beta^2} \bigg)} \Big].
\ee
Then, let us define heat conductivity $\kappa$ in the standard way, 
i.e., as the system response to the applied temperature gradient, under the   condition that the remaining 
currents are equal to zero. It leads to the conclusion that $\kappa$ is of the form as follows:
\be
\kappa = \tkappa + \frac{ \alpha_F T (\alpha_B~\sigma_{FB} - \alpha_F~\sigma_{BB})}{\sigma_{FF}~\sigma_{BB} - \sigma_{FB}^2}
+ \frac{ \alpha_B T (\alpha_F~\sigma_{BF} - \alpha_B~\sigma_{FF})}{\sigma_{FF}~\sigma_{BB} - \sigma_{FB}^2},
\ee
and after some algebra, it reduces to
\be
\kappa = \frac{\tkappa}{1 + \frac{Q^2}{\beta^2 (1- \frac{\alpha^2}{4})(1+g)^2}    \Big[ (1 + \frac{\alpha}{2} g)^2 + (g + \frac{\alpha}{2})^2 - \alpha~(1 + \frac{\alpha}{2}g ) (g + \frac{\alpha}{2} ) \Big]      }.
\ee


\section{Thermoelectric transport coefficients with magnetic field}
In the next step we calculate the DC conductivities of the two-dimensional system with perpendicular magnetic field,
by taking the adequate derivatives from the boundary currents. They are provided as follows:
\ben
\sigma_{(FF)}^{ij} &=& \delta^{ij} \Bigg[
1 + \frac{8 \tQ_{(F)}^2 \Big( \frac{B^2}{r_h^2} + 8 \beta^2 \Big) + 32 B^2 \tQ_{(F)}^2 + B^2 \Big( \frac{B^2}{r_h^2} + 8 \beta^2 \Big)}
{ \Big( \frac{B^2}{r_h^2} + 8 \beta^2 \Big)^2 + 16 ~ B^2 \tQ_{(F)}^2 } \Bigg]\\ \nonumber
&-& \ep^{ij} \Bigg[
\frac{ 8 B \tQ_{(F)} \Big( \frac{B^2}{r_h^2} + 8 \beta^2 \Big) + 32 \tQ_{(F)}^3 B + 8 B^3 \tQ_{(F)}}{ \Big( \frac{B^2}{r_h^2} + 8 \beta^2 \Big)^2 + 16 ~ B^2 \tQ_{(F)}^2 } \Bigg],
\een

\ben
\sigma_{(FB)}^{ij} &=& \sigma_{(BF)}^{ij} =\delta^{ij} \Bigg[
\frac{\alpha}{2} + \frac{8 \tQ_{(F)} \tQ_{(B)} \Big( \frac{B^2}{r_h^2} + 8 \beta^2 \Big) + 16 B^2 \tQ_{(F)} \tQ_{(B)} }{ \Big( \frac{B^2}{r_h^2} + 8 \beta^2 \Big)^2 + 16 ~ B^2 \tQ_{(F)}^2 } \Bigg]\\ \nonumber
&-& \ep^{ij} \Bigg[
\frac{ 4 \tQ_{(B)} B \Big( \frac{B^2}{r_h^2} + 8 \beta^2 \Big) + 32 B \tQ_{(F)}^2 \tQ_{(B)}}{ \Big( \frac{B^2}{r_h^2} + 8 \beta^2 \Big)^2 + 16 ~ B^2 \tQ_{(F)}^2 } \Bigg],
\een

\be
\sigma_{(BB)}^{ij} = \delta^{ij} \Bigg[
1 + \frac{8 \tQ_{(B)}^2 \Big( \frac{B^2}{r_h^2} + 8 \beta^2 \Big)}{ \Big( \frac{B^2}{r_h^2} + 8 \beta^2 \Big)^2 + 16 ~ B^2 \tQ_{(F)}^2 } \Bigg]
- \ep^{ij} \frac{32 B \tQ_{(B)}^2 \tQ_{(F)}}{ \Big( \frac{B^2}{r_h^2} + 8 \beta^2 \Big)^2 + 16 ~ B^2 \tQ_{(F)}^2 }.
\ee 

Next, the thermoelectric conductivities yield
\ben
\alpha_{(F)}^{ij} &=&  16 \pi ~r_h^2~ \delta^{ij} \frac{ 2  \tQ_{(F)} \Big( \frac{B^2}{r_h^2} + 8 \beta^2 \Big) + 4 B^2 \tQ_{(F)}}{ \Big( \frac{B^2}{r_h^2} + 8 \beta^2 \Big)^2 + 16 ~ B^2 \tQ_{(F)}^2 } \\ \nonumber
&-& 16 \pi~ r_h^2 ~\ep^{ij} 
\frac{8 B \tQ_{(F)}^2 + B \Big( \frac{B^2}{r_h^2} + 8 \beta^2 \Big)}{ \Big( \frac{B^2}{r_h^2} + 8 \beta^2 \Big)^2 + 16 ~ B^2 \tQ_{(F)}^2 } + \frac{B}{2T} ~\ep^{ij} ~\Sigma_{(1)},
\een

\ben
\alpha_{(B)}^{ij} &=&  32 \pi ~r_h^2~ \delta^{ij} \frac{   \tQ_{(B)} \Big( \frac{B^2}{r_h^2} + 8 \beta^2 \Big)}{ \Big( \frac{B^2}{r_h^2} + 8 \beta^2 \Big)^2 + 16 ~ B^2 \tQ_{(F)}^2 } \\ \nonumber
&-& 16 \pi~ r_h^2 ~\ep^{ij} 
\frac{8 B \tQ_{(F)} \tQ_{(B)}}{ \Big( \frac{B^2}{r_h^2} + 8 \beta^2 \Big)^2 + 16 ~ B^2 \tQ_{(F)}^2 } - \frac{\alpha~B}{4~ T}~\ep^{ij}~\Sigma_{(1)}.
\een

The thermal conductivity is of the form
\be
\kappa^{ij} = 64~\pi^2~r_h^4~T \Bigg[
\delta^{ij} \frac{2 \Big( \frac{B^2}{r_h^2} + 8 \beta^2 \Big)}{ \Big( \frac{B^2}{r_h^2} + 8 \beta^2 \Big)^2 + 16 ~ B^2 \tQ_{(F)}^2 } 
- \ep^{ij} \frac{8 B \tQ_{(F)}}{ \Big( \frac{B^2}{r_h^2} + 8 \beta^2 \Big)^2 + 16 ~ B^2 \tQ_{(F)}^2 } \Bigg] - \frac{B}{T}~\ep^{ij}~\Sigma_{(a)}.
\ee

In \cite{har07,kim15} it was revealed that the terms proportional to  $\Sigma_{(m)} B/T$,  where $m= 1,~a$, emerged from the contributions of magnetization
currents which stemmed from the two considered $U(1)$-gauge fields. In order to find the DC-conductivities, one ought to subtract them from the expressions in question.
It implies
\ben
\sigma_{(ab)}^{ij} &=& \sigma_{(ab)}^{ij},\\
\alpha_{(F)}^{ij} &=& \alpha_{(F)}^{ij} - \frac{B}{2~T}~\ep^{ij}~\Sigma_{(1)} ,\\
\alpha_{(B)}^{ij} &=& \alpha_{(B)}^{ij} - \frac{\alpha~B}{4~T}~\ep^{ij}~\Sigma_{(1)} ,\\
\kappa^{ij} &=& \kappa^{ij} +  \frac{\ep^{ij}~B}{T}~\Sigma_{(a)},
\een
where $a,~b = F,~B$.
All the above quantities are given by the black brane event horizon data.

\section{Dyonic black hole with momentum relaxation in hidden sector}
To discuss the problem more explicitly, we take into account
the ansatz for static four-dimensional topological black brane with planar symmetry of the form as given by (\ref{sp}).
The gauge fields are given by $ A_t = \tmu ( 1 - \frac{r_h}{r} )$ and $A_y =q_m r_h x,~A_x=- q_m r_h y$
for Maxwell field, while for the other gauge sector we provide the ansatz $
B_t = \tmu_{add} ( 1 - \frac{r_h}{r} ).$
The $R_{xx}$ term of Einstein-gauge scalar field  gravity will reveal that
\be
f(r) = \frac{r^2}{L^2}- \frac{\beta^2}{2} - \frac{m}{r} + \frac{(\tmu^2 + \tmu^2_{add} + \alpha \tmu \tmu_{add} +q_m^2) r_h^2}{4~r^2},
\ee
where $m$ is constant.  One can remark, that we get the additional term which mixes the ordinary and the additional charge parameters.
It can be easily found that the ADM (Arnowitt-Deser-Misner) mass of the black object in question also contains the mixing term of the adequate gauge field parameters
\be
m = \frac{r_h^3}{L^2} - \frac{\beta^2}{2}r_h + \frac{(\tmu^2 + \tmu^2_{add} + \alpha \tmu \tmu_{add} +q_m^2) r_h}{4},
\ee
and the Hawking temperature is provided by
\be
T = \frac{1}{4 \pi}~\bigg[ \frac{3 r_h}{L^2} - \frac{\beta^2}{2 r_h} - \frac{(\tmu^2 + \tmu^2_{add} + \alpha \tmu \tmu_{add} +q_m^2)}{4 r_h} \bigg].
\ee

\section{Kinetic and transport coefficients for the spacetime of dyonic black hole with two interacting gauge fields}
If we denote by $\mu^2 = 1/8 \beta^2 r_h^2$, then the adequate kinetic 
and transport coefficients can be written as follows:
\ben
\sigma_{(FF)}^{ij} {} \\ \nonumber
&= & \delta^{ij} \Bigg[
1 + \frac{8 (\mu \tQ_{(F)} r_h)^2 (B^2 \mu^2 +1) + 32 (\mu \tQ_{(F)} r_h)^2 (\mu B r_h)^2 + (\mu B  r_h)^2 (B^2 \mu^2 +1)}
{(B^2 \mu^2 +1)^2  + 16~(\mu B r_h)^2~(\tQ_{(F)} \mu r_h)^2} \Bigg] \\ \nonumber
&-& \ep^{ij} \Bigg[
\frac{ 8 (\mu B r_h) (\mu \tQ_{(F)} r_h) (B^2 \mu^2 +1) + 32 (\mu \tQ_{(F)} r_h)^3 (\mu B r_h) + 8 (\mu B r_h)^3 (\mu \tQ_{(F)} r_h)}
{(B^2 \mu^2 +1)^2  + 16~(\mu B r_h)^2~(\tQ_{(F)} \mu r_h)^2} \Bigg],
\een

\ben 
\sigma_{(FB)}^{ij} &=& \sigma_{(BF)}^{ij} {}\\ \nonumber
&=&
\delta^{ij} \Bigg[ \frac{\alpha}{2} + \frac{ 8 (\mu \tQ_{(F)} r_h) (\mu \tQ_{(B)} r_h) (B^2 \mu^2 +1) + 16 (\mu B r_h)^2 (\mu \tQ_{(F)} r_h)(\mu \tQ_{(B)} r_h)}
{(B^2 \mu^2 +1)^2  + 16~(\mu B r_h)^2~(\tQ_{(F)} \mu r_h)^2} \Bigg] \\ \nonumber
&-& \ep^{ij} \Bigg[
\frac{4 (\mu \tQ_{(B)} r_h) (\mu B r_h) (B^2 \mu^2 +1) + 32 (\mu B r_h)(\mu \tQ_{(F)} r_h)^2 (\mu \tQ_{(B)} r_h) }
{(B^2 \mu^2 +1)^2  + 16~(\mu B r_h)^2~(\tQ_{(F)} \mu r_h)^2} \Bigg],
\een

\ben
\sigma_{(FF)}^{ij} 
&= & \delta^{ij} \Bigg[ 1 + 
\frac{ 8 (\mu \tQ_{(B)} r_h)^2  (B^2 \mu^2 +1)}{(B^2 \mu^2 +1)^2  + 16~(\mu B r_h)^2~(\tQ_{(F)} \mu r_h)^2} \Bigg] \\ \nonumber
&-& \ep^{ij}~
\frac{32 B~(\mu \tQ_{(B)} r_h)^2 (\mu \tQ_{(F)} r_h)}{(B^2 \mu^2 +1)^2  + 16~(\mu B r_h)^2~(\tQ_{(F)} \mu r_h)^2},
\een

\ben
\alpha_{(F)}^{ij} &=&  16 \pi ~r_h^2~ \delta^{ij} \frac{ 2\mu r_h \Big[ (\mu \tQ_{(F)} r_h)(B^2 \mu^2 +1) + 4 (\mu B r_h)^2 (\mu \tQ_{(F)} r_h) \Big]}
{(B^2 \mu^2 +1)^2  + 16~(\mu B r_h)^2~(\tQ_{(F)} \mu r_h)^2} \\ \nonumber
&-& 16 \pi~ r_h^2 ~\ep^{ij} 
\frac{\mu r_h \Big[ 8 (\mu B r_h)(\tQ_{(F)} \mu r_h)^2 + (\mu B r_h)(B^2 \mu^2 +1) \Big]}
{(B^2 \mu^2 +1)^2  + 16~(\mu B r_h)^2~(\tQ_{(F)} \mu r_h)^2},
\een

\ben
\alpha_{(B)}^{ij} &=&  32 \pi ~r_h^2~ \delta^{ij} \frac{\mu r_h (\mu \tQ_{(B)} r_h)(B^2 \mu^2 +1)}{(B^2 \mu^2 +1)^2  + 16~(\mu B r_h)^2~(\tQ_{(F)} \mu r_h)^2} \\ \nonumber
&-& 16 \pi~ r_h^2 ~\ep^{ij} 
\frac{8 \mu^2 r_h^2 B (\tQ_{(F)} \mu r_h)(\tQ_{(B)} \mu r_h)}{(B^2 \mu^2 +1)^2  + 16~(\mu B r_h)^2~(\tQ_{(F)} \mu r_h)^2},
\een

\ben
\kappa^{ij} = 64~\pi^2~r_h^4~T \Bigg[
\delta^{ij} \frac{2 \mu^2 r_h^2 (B^2 \mu^2 +1)}{(B^2 \mu^2 +1)^2  + 16~(\mu B r_h)^2~(\tQ_{(F)} \mu r_h)^2} \\ \nonumber
-\ep^{ij} \frac{8 \mu^2 r_h^2 (\mu B r_h)(\tQ_{(F)} \mu r_h)}{(B^2 \mu^2 +1)^2  + 16~(\mu B r_h)^2~(\tQ_{(F)} \mu r_h)^2} \Bigg],
\een
where in the context of the previous section one has that
\be
\tQ_{(F)} = \Big(\tmu + \frac{\alpha}{2} \tmu_{add} \Big) r_h, \qquad
\tQ_{(B)} = \Big( \tmu_{add} + \frac{\alpha}{2} \tmu \Big) r_h, \qquad B= q_m r_h.
\ee
It has to be noted again that the parameter $\mu$ plays a role of the mobility in real materials. This interpretation
is supported not only by its place in the above formulas, but also the interpretation
of $\beta$ leading to the  momentum relaxation on a gravity side. 

One can envisage that the effect of momentum relaxation $\beta$, mobility $\mu$, magnetic field $B$ and $\alpha$-coupling constant is not easily observed due to the fact that
$r_h$ is rather complicated function of $\tmu,~\tmu_{add},~q_m$ and depends moreover 
on the coupling constant between both  sectors. However,
the knowledge of the above kinetic coefficients
allows us to calculate the respective transport parameters, the resistivity 
tensor $\rho^{ij}$ which components are given by the
inverse of the conductivity matrix $\sigma$  and the Nernst and Seebeck parameters. 
The latter coefficient $S\equiv S^{xx}$ is defined as
a longitudinal voltage (in the direction of temperature gradient) induced by the 
unit temperature gradient under the condition that no charge current flows.  
The Seebeck and Nernst transport coefficients are given by the adequate elements of the matrix
\be
S^{ij}=(\sigma^{-1})^{il}\alpha_l^j.
\ee

\section{Confrontation with experiments}
\begin{figure} 
\includegraphics[width=0.45\linewidth]{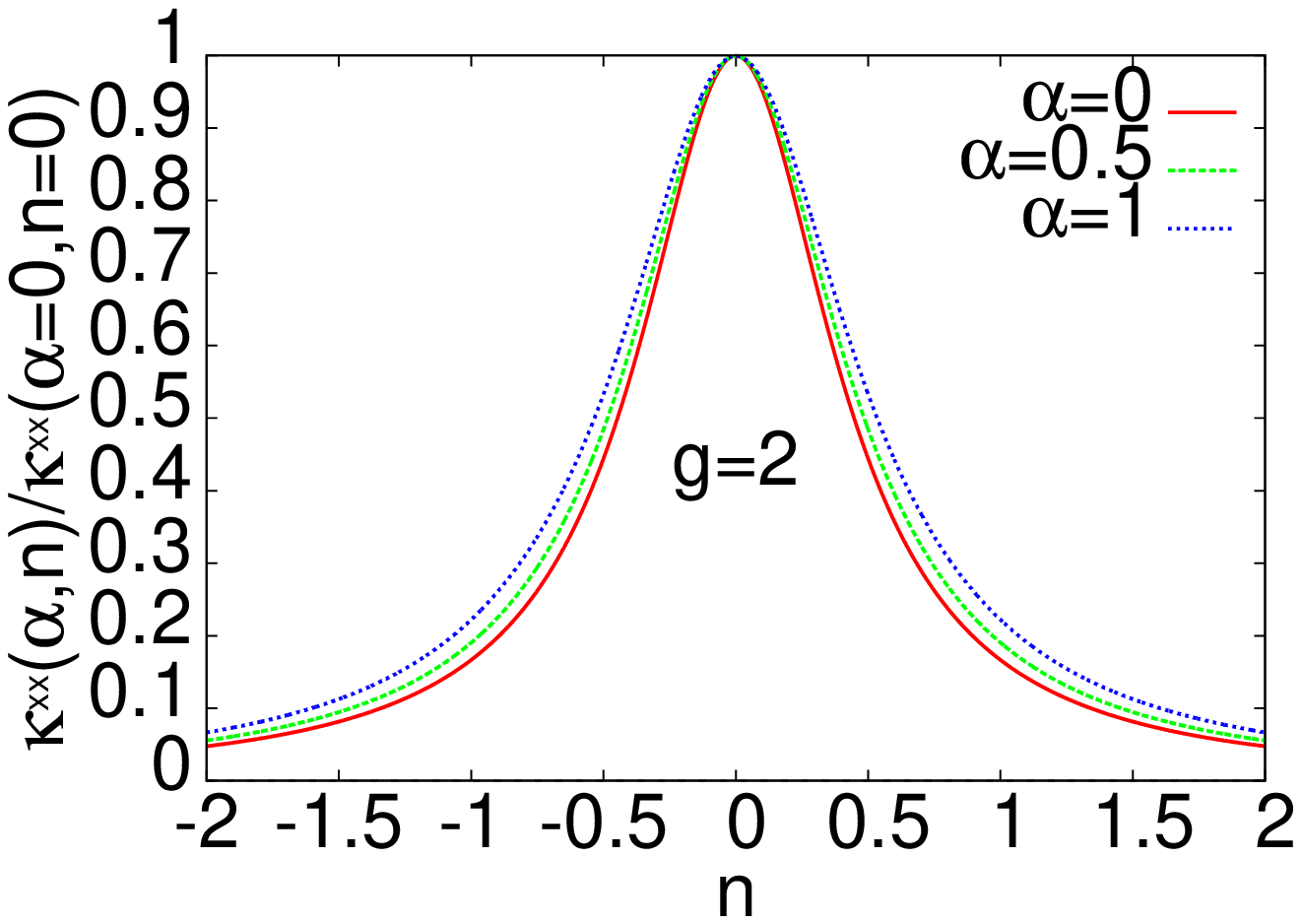} \hspace{1.3cm}
\includegraphics[width=0.45\linewidth]{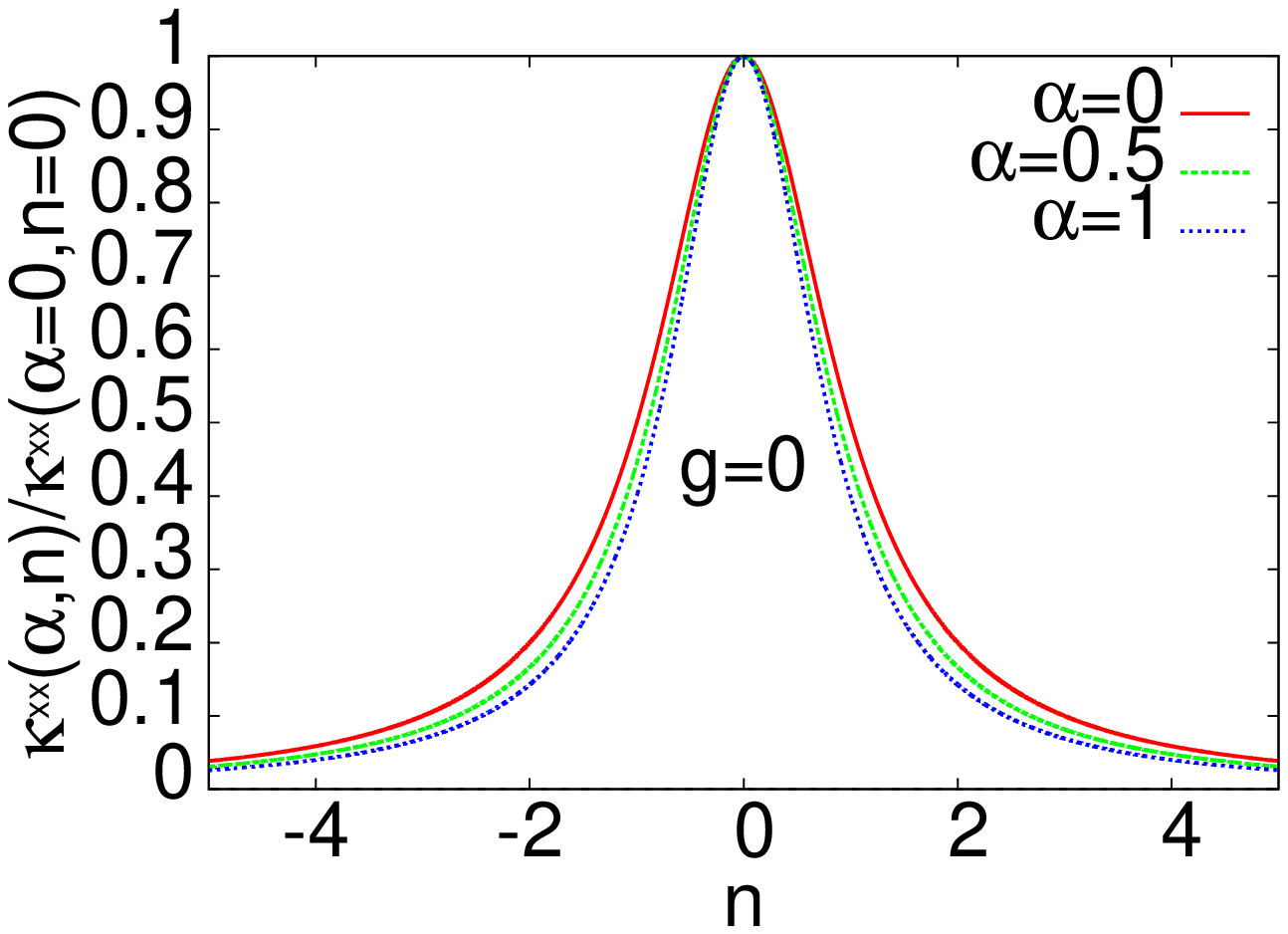}
\caption{ {(color online)(left panel) Charge carrier  dependence of the  thermal conductivity $\kappa^{xx}$,
normalized to its $\alpha=0$ value at $n=0$, obtained for the magnetic field $B=0$, 
mobility $\mu=0.5$, $g=2$   and a few values of $\alpha$.  (right panel) 
 The same dependence, except that $g=0$. Note, the  $g$ dependent change of the width of curves
for various values of $\alpha$ parameters.
}}
\label{fig6}
\end{figure}

Transport coefficients of graphene have been experimentally measured
and theoretically analyzed in a number of works (for review see, e.g., \cite{graph-rev1,graph-rev2}). 
Also there exist
a number of papers using holographic approach \cite{har07,crossno2016}. {In the recent paper \cite{seo17} 
thermal conductivity has been measured and analyzed by means of holographic approach within 
 the two current model. Two currents can be envisaged as that of electrons and holes
present in the system with the Fermi energy tuned to coincide with the Dirac point.
The model \cite{seo17} neglects possible excitonic interactions between the charges and corresponds  to  
$\alpha=0$. In the action (\ref{sgrav}) we have considered two fields leading to the two interacting currents.

We thus start to analyze the effect of the coupling $\alpha$  
between the currents on the charge dependence of 
$\kappa^{xx}$ in a model without magnetic field. It is illustrated in Fig.\ref{fig6}, where we show the
dependence of $\kappa^{xx}$ on charge concentration $n$ ($Q=en$) for three values of $\alpha$ and for $g=2$
in the left panel and $g=0$ in the right panel. Both figures refer to the sample with modest mobility $\mu=0.5$. 
The effect is rather small, but the increase of $\alpha$ leads to a slight increase  of the width of the  normalized
thermal conductivity for the model with $g=2$, while the decrease of the width is observed for $g=0$. This shows 
that the very precise agreement of the calculations with experimental data may require the use of the coupling between
these two currents. In all calculations we assume that $r_h=1$ and $T=1$.}

\begin{figure} 
\includegraphics[width=0.45\linewidth]{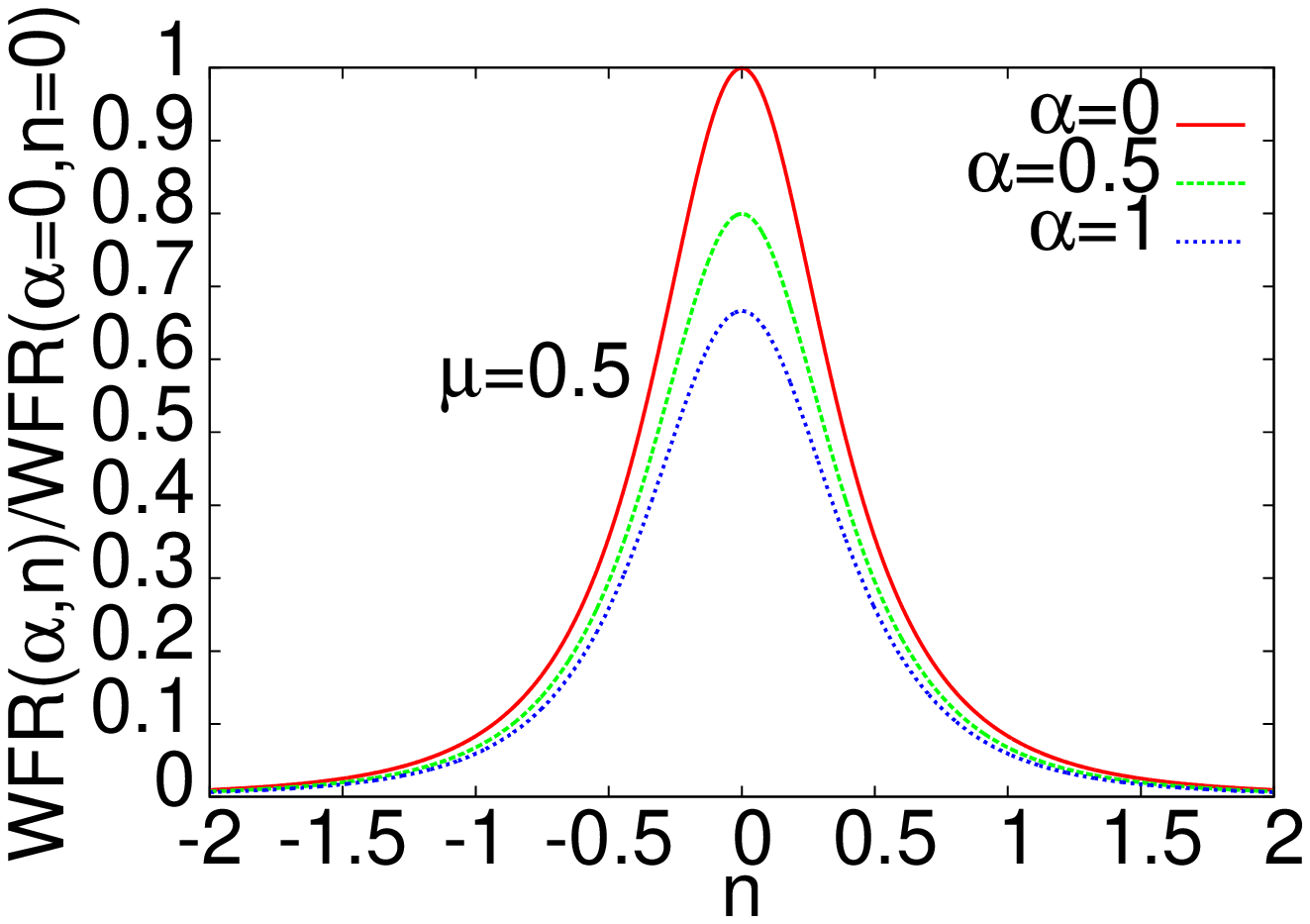} \hspace{1.3cm}
\includegraphics[width=0.45\linewidth]{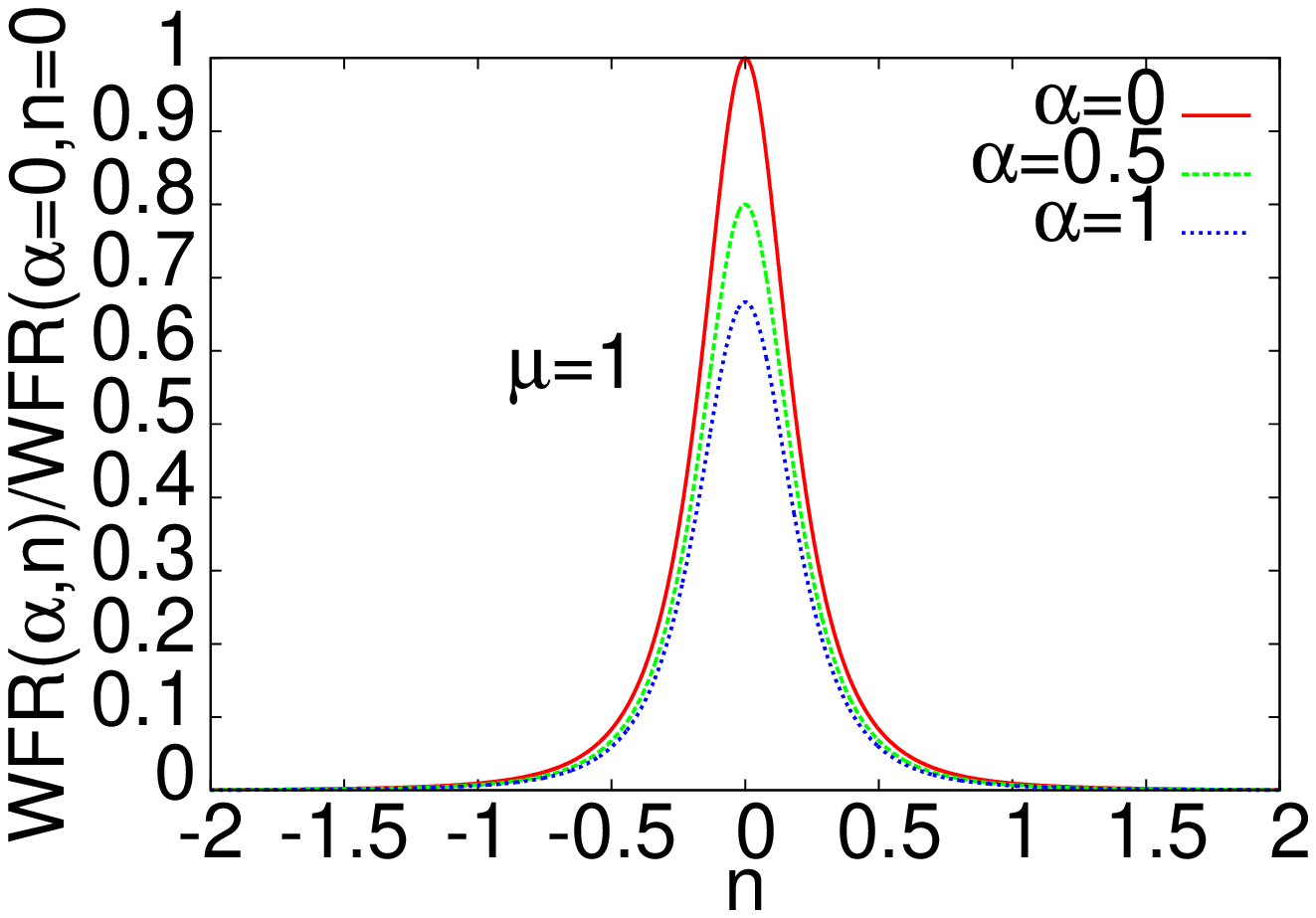}
\caption{ {(color online)(left panel) Charge carrier  dependence of the  Wiedemann-Franz ratio (WFR)
normalized to its $\alpha=0$ value at $n=0$ obtained for $B=0$, mobility $\mu=0.5$, $g=2$  
and a few values of $\alpha$.  (right panel) 
 The same dependence except that $\mu=1$. Note the change of height roughly 
independent of the mobility and $\mu$ dependent change of width at half maximum.
}}
\label{fig5}
\end{figure}

{As next step of our analysis of the effect of $\alpha$ on transport properties of graphene we 
show in Fig.\ref{fig5} the dependence of the Wiedemann - Franz ratio (WFR) defined as 
\be
WFR=W^{xx}=\kappa^{xx}/(\sigma^{xx}T),
\label{wfxx}
\ee  
where $\sigma^{xx}=\sum_{a,b}^{F,B}\sigma^{xx}_{ab}$.
The effect is related to the change of the the width of curves, as well as, their heights. Again the 
precise analysis of the dependence of $WFR$ on $n$ can be achieved by the appropriate use of
both parameters referring to the currents, namely $g$ and $\alpha$. Generally, $WFR$ diminishes with
increase of $\alpha$ for all values of the charge density. This change can be attributed to the increase of 
conductivity or the decrease of resistivity. The latter quantity is shown in the right panel of Fig.\ref{fig5}. 
}

\begin{figure} 
\includegraphics[width=0.45\linewidth]{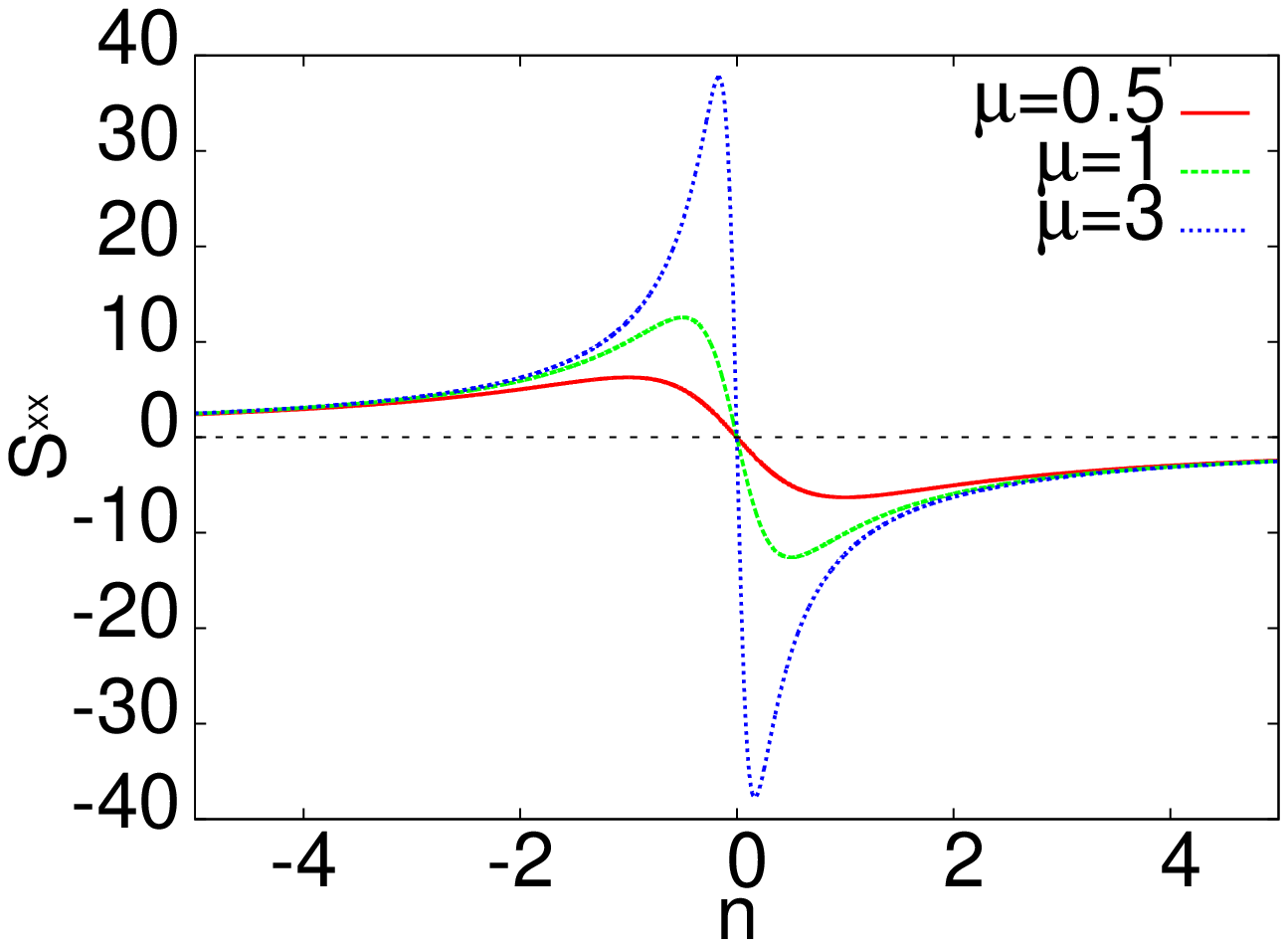} \hspace{1.3cm}
\includegraphics[width=0.45\linewidth]{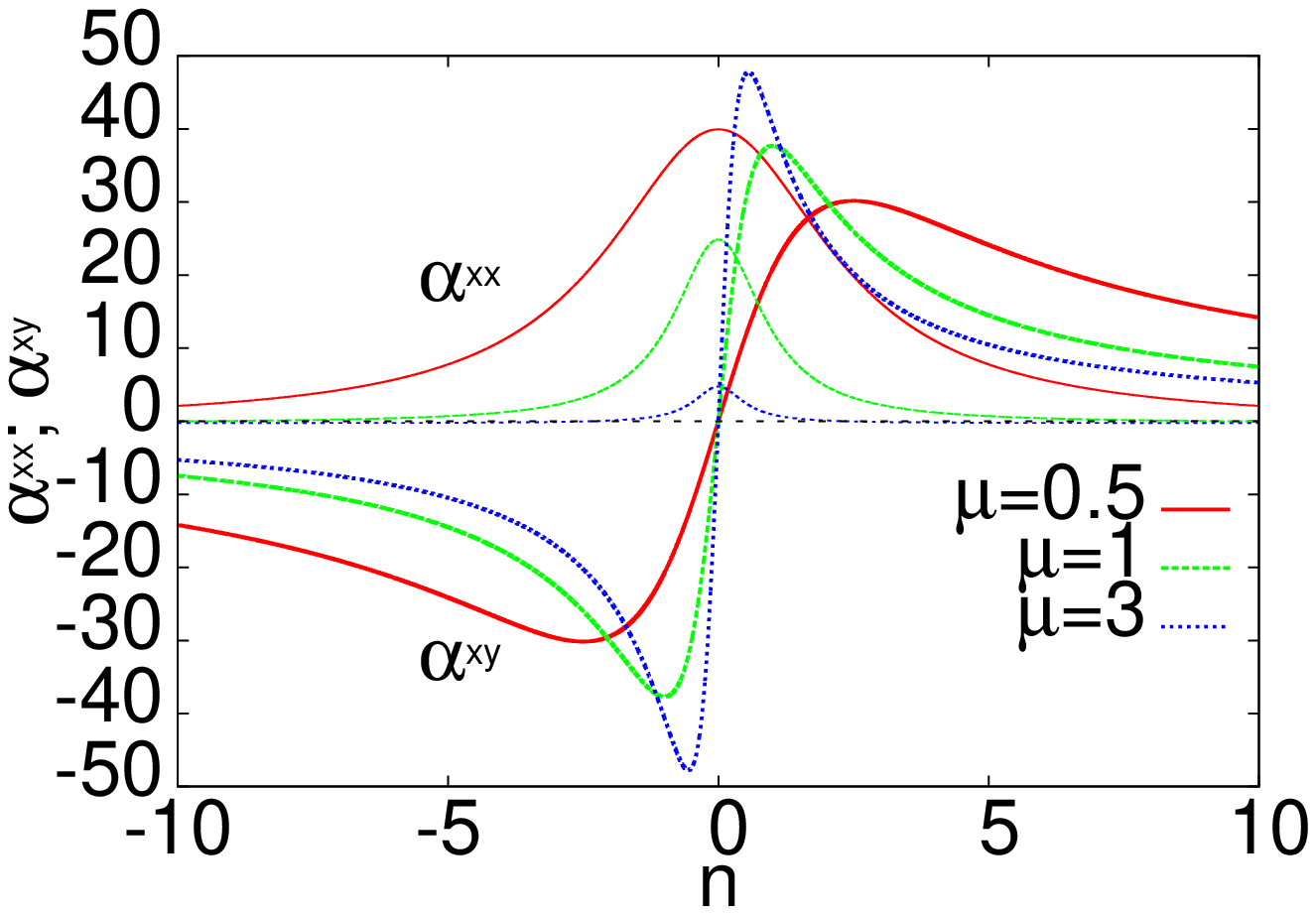}
\caption{ {(color online) (left panel) Charge carrier  dependence of the  Seebeck  coefficient $S=S^{xx}$ 
obtained for $B=0$  and a few values of mobility $\mu$, the parameter which on the
gravity side is related to the momentum dissipation  $\beta$. (right panel) 
The kinetic coefficients $\alpha^{xx}$ and $\alpha^{xy}$ $vs.~n$ obtained for
magnetic field $B=1$  and a few values of $\mu$.
 The coefficient $\alpha^{xy}$ has been shifted upwards by the constant value 50. }}
\label{fig3}
\end{figure}

{In the left panel of Fig.\ref{fig3} we show the 
dependence of the Seebeck coefficient $S^{xx}$ on the charge concentration $n$ for three systems characterized
by different mobilities $\mu=0.5,1$ and $3$. With the increase of mobility $S^{xx}$ gets larger value 
and its maximum shifts towards smaller carrier concentration. The Seebeck coefficient 
has been measured in \cite{ghahari2016} as a function of gate voltage applied to the graphene sheet. 
To see the relevance of our calculations it has to be recalled that the charge concentration in graphene 
can be changed by the external gate voltage. The detailed relation between $n$ and the gate voltage  is 
unknown but is typically of linear character.  
}
The dependence of $S$ on the gate voltage measured for different temperatures \cite{ghahari2016} and shown in
Fig.3 of that paper nicely agrees with our calculations 
as presented in Fig.\ref{fig3} (left panel). In the figure we plot the Seebeck coefficient 
 for a few values of the mobility parameter $\mu$. 
The authors of the experiment suggest that the interaction with the optical phonons
is responsible for the observed changes of $S$ with temperature. As visible in the discussed figure 
we observe completely analogous changes with the 
mobility of the sample in question. This is sensible as  in the ultra-pure graphene 
studied in \cite{ghahari2016} the increased 
interaction with phonons  reduces the mobility of the system at higher temperatures.     
{The very good agreement between the experimentally measured data and our calculations 
can be interpreted in favor of the holographic approach being able to describe real 
systems studied in the lab.}

Similarly, very good agreement with the experimentally determined dependence
of the coefficients $\alpha^{xx}$ and $\alpha^{xy}$ on the carrier concentration 
is observed between our data, shown in the right panel of the Fig.\ref{fig3}, and the dependence
plotted in the Fig.4 of the paper \cite{checkelsky2009}. However, to get the agreement with the
experimental dependence of  $\alpha^{xy}$ we have to shift it vertically by the
constant value $50$. This is probably related to the fact that experiment has been performed at 
high  magnetic fields ($B=7T$ and $14T$). At such values of the field the spectrum becomes
quantized and the occupied Landau level appears at the Dirac point \cite{graph-rev1,graph-rev2}. 
We have not taken into account  this effect in our holographic approach \cite{davis2008,jokela2011} 
and the above shift corrects for it. 

\begin{figure} 
\includegraphics[width=0.45\linewidth]{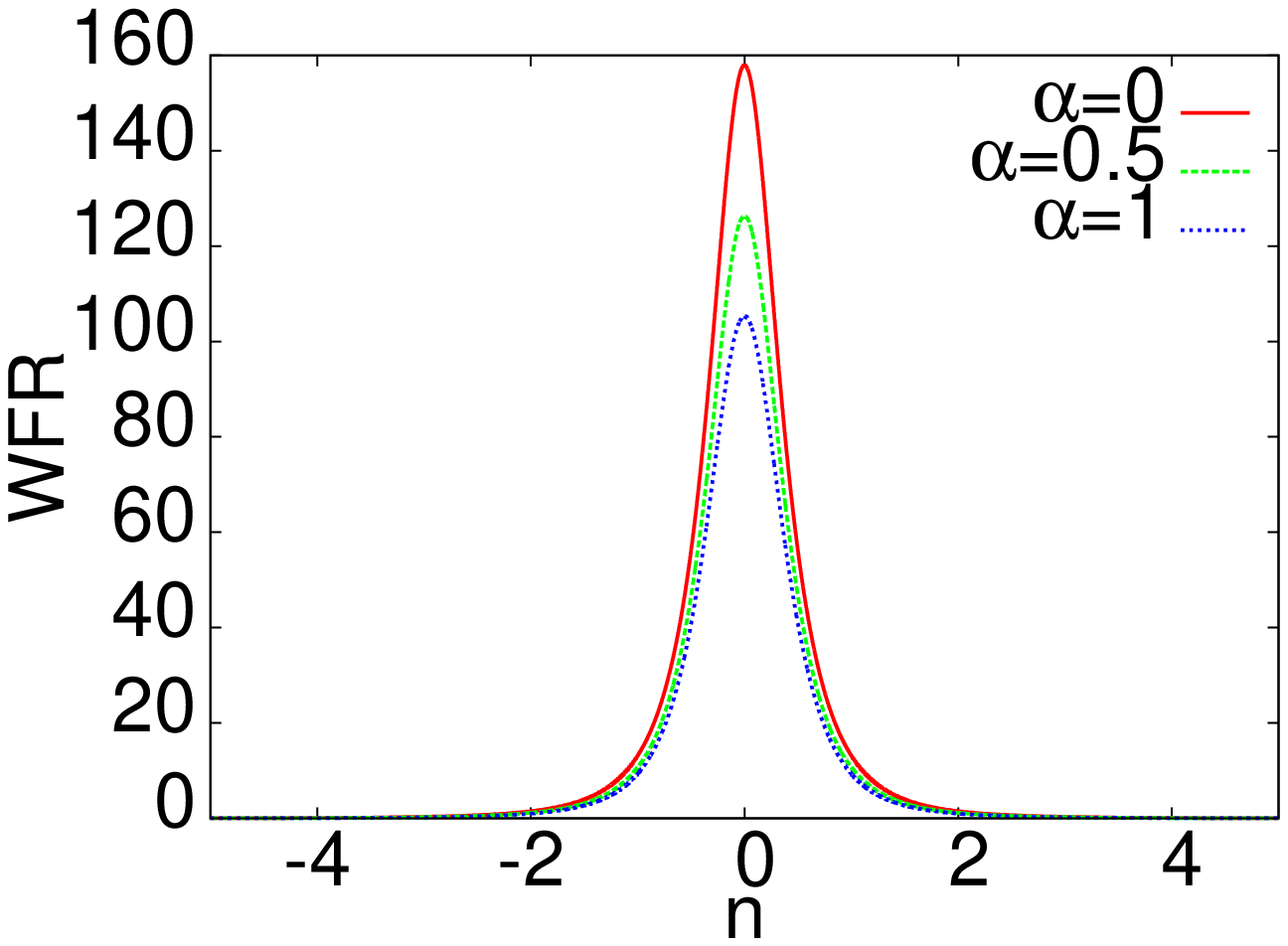} \hspace{1.3cm}
\includegraphics[width=0.45\linewidth]{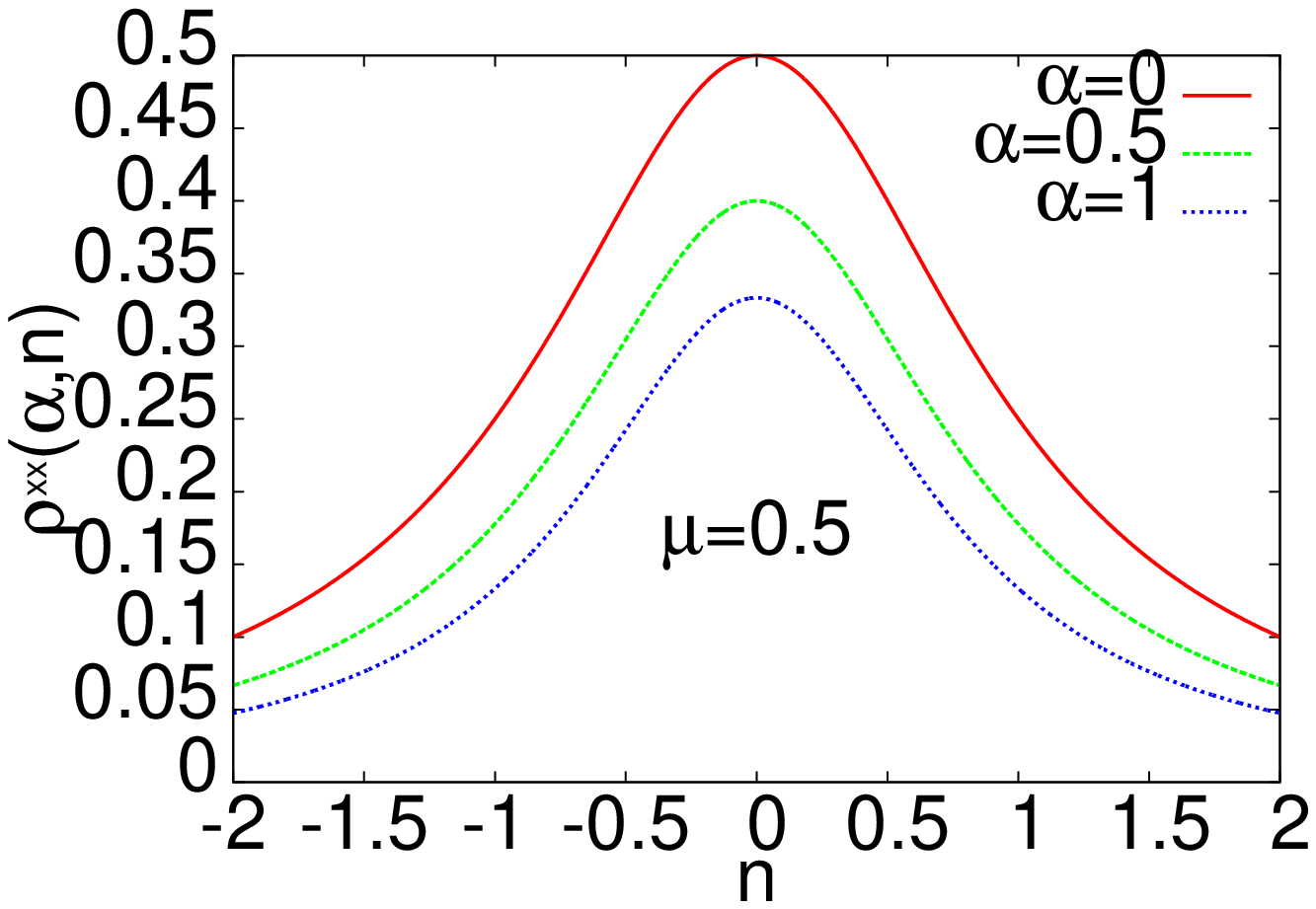}
\caption{ {(color online) (left panel) Charge carrier  dependence of the  Wiedemann-Franz ratio (WFR)
obtained for $B=0$, mobility $\mu=0.5$, $g=2$  and a few values of $\alpha$.  (right panel) 
Similar dependence of the diagonal resistivity $\rho^{xx}$ and the same system parameters.
}}
\label{fig4}
\end{figure}

Finally we comment on the $\alpha$ effect on the diagonal resistivity  
and the Wiedemann-Franz ratio. The charge dependence of these two transport parameters are displayed
in Fig.\ref{fig4}. Increase of $\alpha$ leads to decrease of both $\rho^{xx}$ and $WFR=W^{xx}$. 
Again the effect is not very big but well visible and amounts to change of the maximum value of $W^{xx}$
by 20\% if the coupling $\alpha$ changes from 0 to 0.5. 
  
{It has to be reminded that all transport coefficients of graphene become a two by two matrices if
the  magnetic field $B$ perpendicular to the layer is applied. The important parameter entering all
transport coefficients together with $B$ is the effective mobility $\mu$ related to the 
holographic parameter $\beta$  responsible for the dissipation of momentum. It is important to notice
that the diagonal transport coefficients take on finite values even at zero charge concentration. However,
to have non-zero also the off-diagonal elements one has to assume finite values of the charge density.
Here we shall assume $n=0.1$. With this value of charge density we are close enough to the particle - hole
symmetry point and may analyze the whole matrix of kinetic and transport coefficients. We start with 
Seebeck $S^{xx}$ and Nernst $S_{xy}$ effects.}

\begin{figure} 
\includegraphics[width=0.45\linewidth]{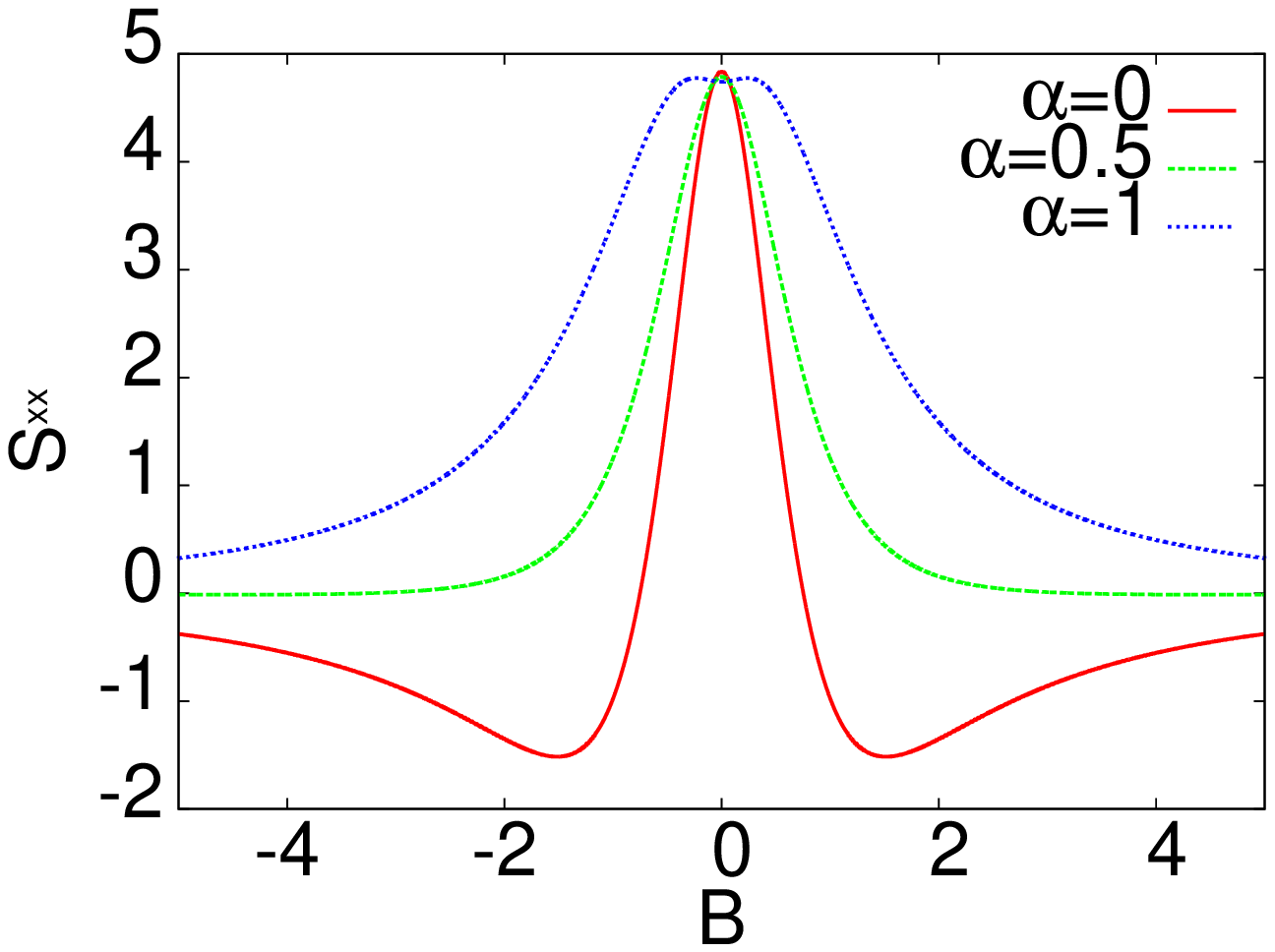} \hspace{1.3cm}
\includegraphics[width=0.45\linewidth]{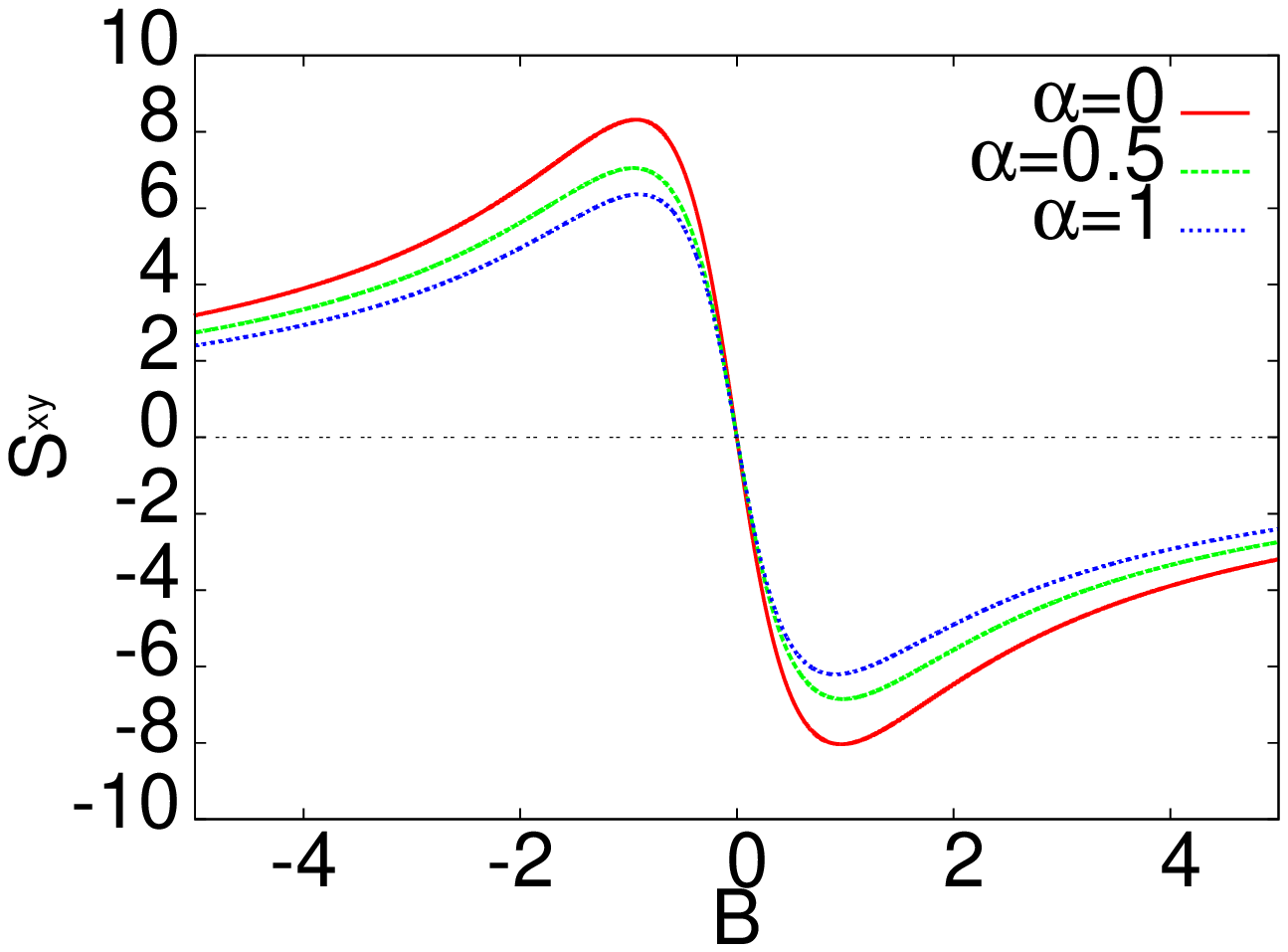}
\caption{{ (color online) The Seebeck $S^{xx}$ (left panel) and Nernst $S^{xy}$ (right panel) 
coefficients calculated for $n=0.1$, $\mu=1$, $g=2$ and  a few values of the coupling $\alpha$  
as functions of magnetic field $B$.
}}
\label{fig1}
\end{figure}
\begin{figure} 
\includegraphics[width=0.45\linewidth]{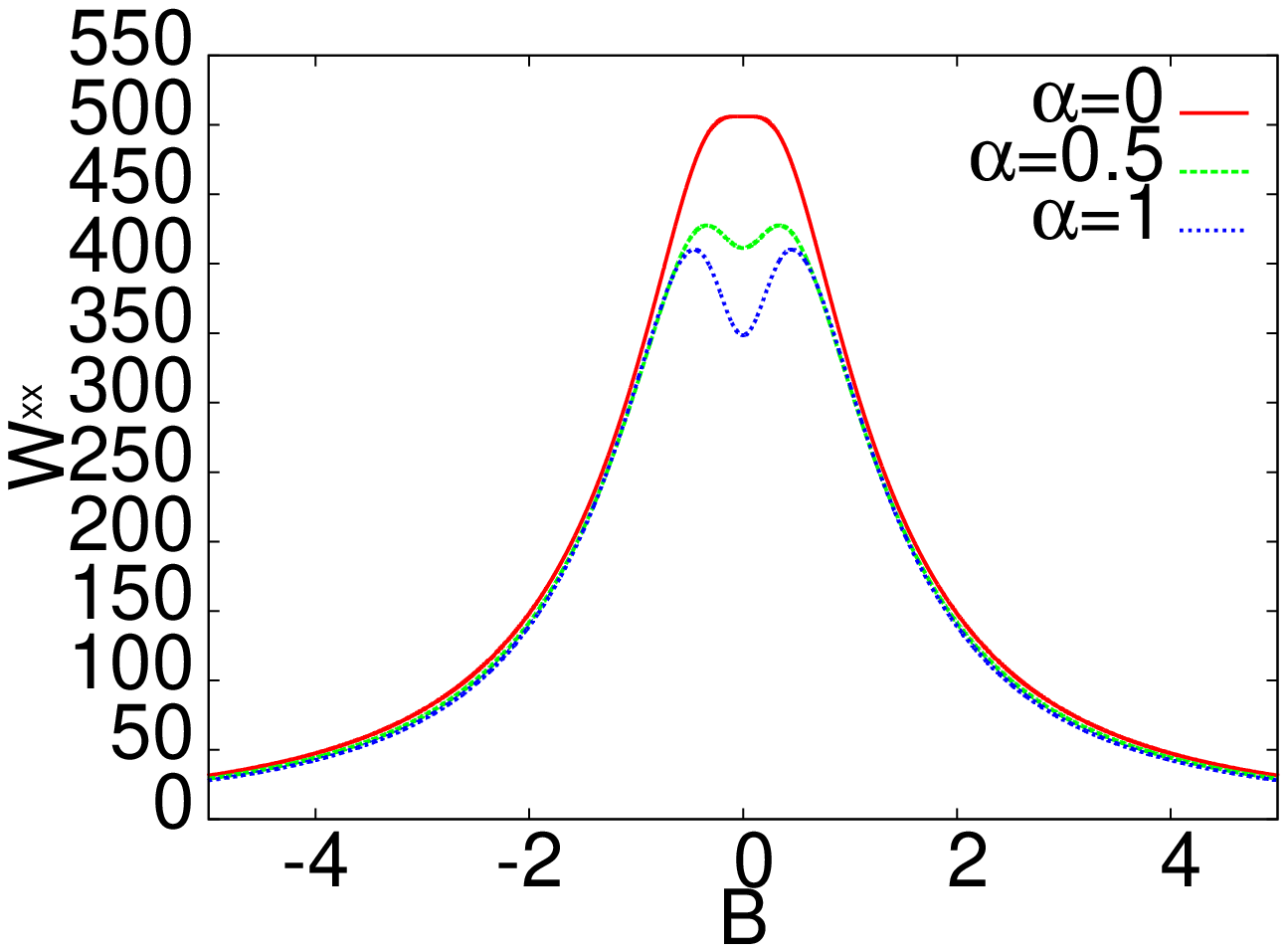} \hspace{1.3cm}
\includegraphics[width=0.45\linewidth]{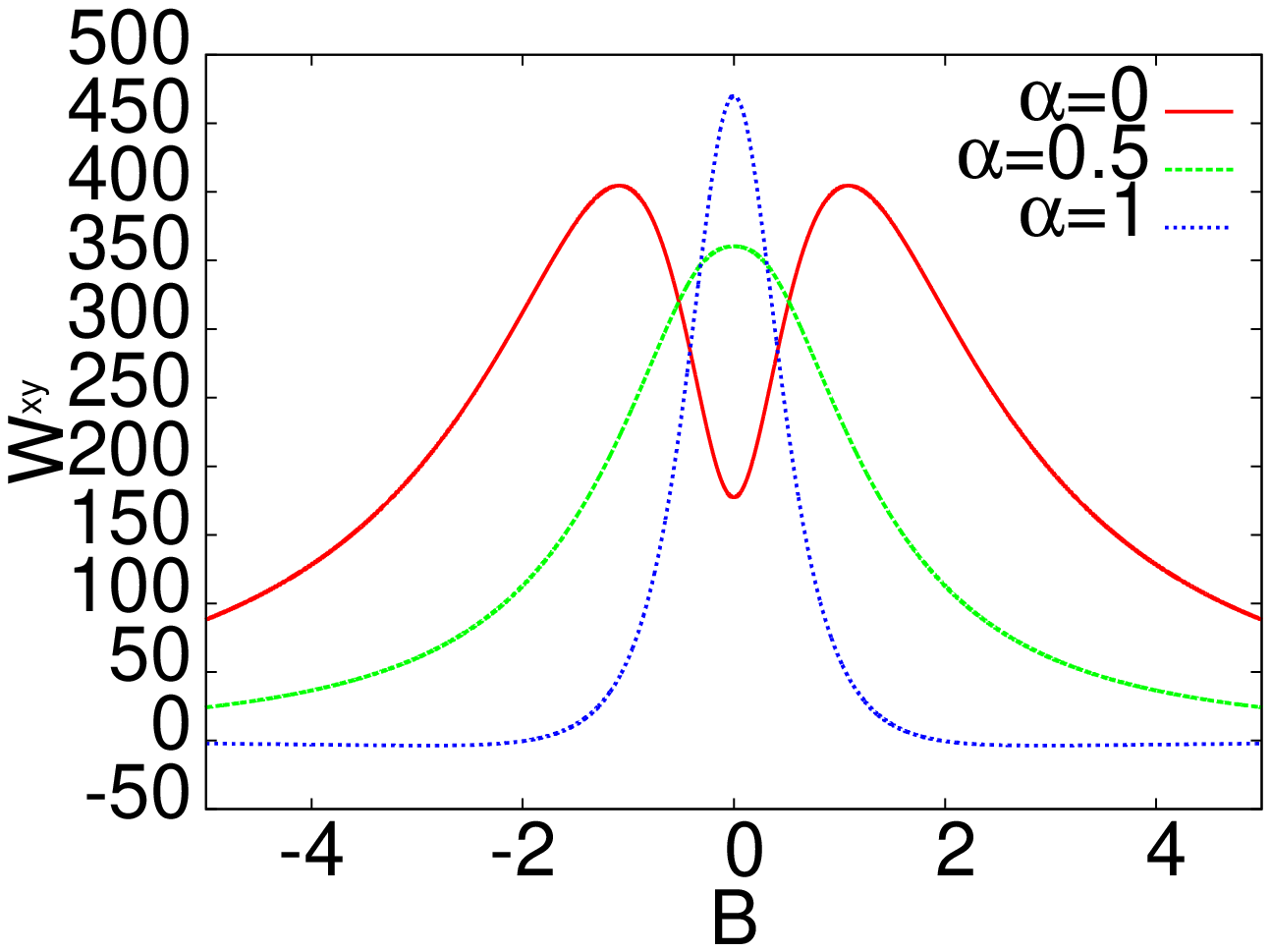}
\caption{{(color online) Magnetic field  dependence of the diagonal $WFR$  (left panel) and
off-diagonal  $WFR^{xy}$   Wiedemann - Franz ratio (right panel) 
as function of magnetic field $B$ and for a few values of the coupling $\alpha$.
Other parameters are set to $\mu=1$ and $g=2$. }}
\label{fig2}
\end{figure}

{Fig.\ref{fig1} illustrates the magnetic field dependence of the Seebeck and
Nernst coefficients for moderate value of the mobility $\mu=1$ and for the current
mixing parameter $g=2$ (this is close  to the value used to describe charge dependence of
thermal conductivity in graphene \cite{seo17}). Again we pay special attention to the 
effect of $\alpha$ on the studied dependencies. It is especially large on the $S^{xx}(B)$
with spectacular change of shape: from the curve with  two minima and a maximum for $B=0$
observed for $\alpha=0$ to the curve with a minimum at $B=0$ and two small maxima for larger absolute
value of the magnetic field. The Nernst coefficient $S^{xy}$ is an anti-symmetric function of $B$ while
$S^{xx}$ is symmetric in $B$. 
}

{Typically one measures the Wiedemann-Franz ratio for a system at zero or constant magnetic field
varying the charge density. Here we propose the generalization of this parameter in two directions. First,
we define both diagonal and off-diagonal parts and second we study it as a function 
of magnetic field. While $W^{xx}$ is defined in Eq. (\ref{wfxx}), we define $W^{xy}$ in the
simplest possible way as
\be
W^{xy}=\frac{\kappa^{xy}}{T\sigma^{xy}}.
\ee
We are not aware
of any experimental work on graphene studying systematically  these parameters as functions of the magnetic
field for constant charge density and propose their measurements as a possible check of our theory and holographic
analysis of transport in graphene.  Such measurements would provide an important hint towards holographic
modeling of transport in strongly interacting systems. Our predictions of the magnetic field 
dependence of $W^{xx}$ and $W^{xy}$ are shown in Fig.\ref{fig2}. 
}

\subsection{The Hall angle}
In this subsection we shall elaborate the influence of $\alpha$-coupling constant of the two sectors in question on the hall angle.
To commence with, 
let us define the Hall angle, by the ratio of the electric conductivities, in the form provided by
\be 
\tan \theta = \frac{\sigma^{xx}}{\mid \sigma^{xy} \mid} = \frac{E}{F},
\ee
where we have denoted
\ben
\sigma^{xx} &=& \sigma_{(FF)}^{xx} + 2 \sigma_{(FB)}^{xx} + \sigma_{(BB)}^{xx},\\
\sigma^{xy} &=& \sigma_{(FF)}^{xy} + 2 \sigma_{(FB)}^{xy} + \sigma_{(BB)}^{xy},
\een
The exact forms of $\sigma_{(a b)}^{ij}$ lead to the following expressions for $E$ and $F$:
\ben
E &=& (2+\alpha)[(B^2 + 8 \beta^2 r_h^2) + 16 B^2 \tmu^2~ r_h^6] + 8 \tmu~ r_h^4 ~(B^2 + 8 \beta^2 r_h^2) \Big(1 + \frac{\alpha}{2}\Big)^2 (1+g)^2 \\ \nonumber
&+& 32 B^2 \tmu^2~ r_h^6 ~ \Big(1 + \frac{\alpha}{2}\Big)(1+g)
+ B^2~ r_h^2~ (B^2 + 8 \beta^2 r_h^2),
\een
and
\ben
F &=& 32 B \tmu^3~ r_h^7 ~\Big(1 + \frac{\alpha}{2}g \Big) \Big(1 + \frac{\alpha}{2}\Big)^2(1 + g)^2 + 8B^2 \tmu~ r_h^5 ~\Big(1 + \frac{\alpha}{2}g \Big) \\ \nonumber
&+& 8 b \tmu ~r_h^3 ~\Big(1 + \frac{\alpha}{2}\Big)(1 + g)(B^2 + 8 \beta^2 r_h^2).
\een
The explicit value of the charge connected with Maxwell field is given by $Q_{(F)} = \tmu~r_h$. 
On the other hand, for the radius of black brane one obtains the relation
\be
r_{h~(1,2)} = \frac{16 \pi T \pm \sqrt{(16 \pi T)^2 + 48 (2 \beta^2 + \tmu_{all}^2 + q_m^2)}}{24},
\ee
where $\tmu_{all} = \tmu^2 + \tmu_{add}^2 + \alpha \tmu \tmu_{add}$.
Thus $r_h$ is roughly proportional to the Hawking temperature. 
From the above expression, it can be seen that in the limit of high temperature, when $\beta $ tends to zero, 
one gets that $\tan \theta $ increases when $B$ and $\beta$ increase.
Moreover for the limit in question we obtain the proportionality of the Hall angle to the inverse 
of the adequate power of the temperature
\be
\tan \theta = c_0+  \frac{c_1}{T} + \frac{c_2}{T^3} + \cO(1/T^7),
\ee
where the coefficients are provided by
\be
c_1 = \frac{B~(2+ \alpha)}{2 \tmu~(1 + \frac{\alpha}{2} g) (1 +\frac{\alpha}{2})^2(1+g)^2}, \qquad
c_2 = \frac{B}{4 \tmu~(1 + \frac{\alpha}{2} g)}.
\ee
The close inspection of the above coefficients reveals, that for a constant value of magnetic and electric field $\tmu$, ~$\alpha > 0$ and for $g=0.3$ the dominant role plays the term proportional to $1/T^3$.
The bigger value of $\alpha$-coupling constant (and/or $g$) one considers, 
 the greater $c_2$ is, in comparison to $c_1$.


\section{Summary and conclusions}

 We have studied thermoelectric  transport properties of graphene assuming that close
to the Dirac point the carriers are strongly interacting and thus the gauge-gravity duality
is applicable. We consider Hall effect geometry with the magnetic field perpendicular to
the graphene plane and with the electric field and temperature gradients in the plane but 
being perpendicular to each other. The calculation of the DC-transport coefficients is
facilitated by the introduction of the axionic field $\beta$ which on the condensed matter side 
provides momentum relaxation mechanism and, as our calculations show, is related
to the mobility of the material.  
The second sector of $U(1)$-gauge field taken into account in the action affects
the kinetic and transport coefficients {\it via} the parameters $g$ and $\alpha$.  

{Having in mind the reference \cite{seo17}, our model predicts that the increase of $\alpha$-coupling 
constant value leads to the increase of the width of normalized thermal conductivity with
$g=2$. On the contrary, when $g=0$, the effect is quite opposite, i.e., one obtains the decrease of the width.
The dependence of $\alpha$-coupling constant on the Wiedemann-Franz ratio (WFR) is related 
to the changes of the width of curves and their heights. The general tendency envisaged in the fact that WFR
diminishes as $\alpha$-coupling constant increases. The aforementioned dependence is valid for all charge densities. }

{Based on the model in question we plot the dependence of the Seebeck coefficient on the charge concentration, for the different values of mobilities $\mu$. The mobility
increase causes that $S^{xx}$ reaches larger values and its maximum is shifted towards the values of small carrier concentrations. One receives a very good agreement with the experimental data.
The same is true for $\alpha^{xx}$ and $\alpha^{xy}$ coefficients. }

{As far as the charge dependence of the diagonal resistivity and the Wiedemann-Franz ratio 
on $\alpha$-coupling constant, we reveal that the increase of the coupling constant of two gauge fields causes 
the decrease of both $\rho^{xx}$ and $W^{xx}$. We also examine the influence of magnetic field on the Seebeck and Nerst coefficients, paying special attention to the
$\alpha$-coupling constant effects on the aforementioned phenomena. One finds that the influence is large for $S^{xx}$, changing the shape of the curve, from the curve with two minima and a maximum
(for $B=0,~\alpha=0$) to the curve with a minimum at $B=0$ and two small maxima for larger absolute values of magnetic field. To our knowledge, this is the new effect, which has not been observed yet.
Perhaps future experiments may verify our theoretical predictions.}

It also turns out that $\alpha$ influences the Hall angle, causes its increase when magnetic field and 
 $\beta$ increase. In the high temperature regime we observe
that $\tan \theta = c_0 + c_1/T + c_2/T^3 + \cO(1/T^7)$.

However, due to the fact that $\alpha$ modifies the pre-factors
only its experimental detection in such measurements will be very hard, if possible at all. 
The possible exception is provided by the magnetic field dependence of the Seebeck coefficient
 $S^{xx}$ and the diagonal Wiedemann-Franz ratio $W^{xx}$. The situation
might change in the geometry with the in-plane magnetic field. 
It has to be stressed that our results on the density dependence 
of the thermoelectric coefficients $\alpha^{xx}$ and $\alpha^{xy}$
and the Seebeck coefficient $S^{xx}$ nicely agree with 
the experimental data \cite{ghahari2016,checkelsky2009}.

\subsection{Dark matter interpretation}
On the other side, the hope is that experimental  studies
of various condensed systems allow for checks of the approach and eventually contribute to better
understanding of gravity itself. In particular the long standing
problem on the gravity side is the direct observation of the {\it dark matter}. This elusive component
of the Universe is expected to be responsible for more than five times  of the mass in the Universe 
as visible one. The problem is thus serious and worth studying
in view of the latest astronomical observations, proposed future investigations and
negative or non-conclusive results of the present direct experiments 
\cite{agnes2015,til15,massey15a,massey15b,afa09,gni08,suz15,mir09,red13,reg15,ali15,foo15,foo15a,bra14,ful15,lop14,nak12,
nak15aa,mun16}  aiming at its detection. 
There has been some efforts 
to look again into the old astrophysical observations like supernova 1987A data and to try to reinterpret them
taking into account the existence of {\it dark radiation} (the dark photon) \cite{cha17}, as well as, to find the strong constraints on emission of {\it dark photons} 
from stars \cite{pos13} and on the coupling of {\it dark matter} coming from 
light particle production in hot star cores and their effects on star cooling \cite{har17}.
The aforementioned studies are also important in the context of the new rival precession of cosmic microwave background measurements, delivered
by Dark Energy Survey (equipped with 570-megapixel camera, able to capture the digital imagines of galaxies at 8 billion light years distances) which supports the view that 
{\it dark matter} and {\it dark energy} make up most of our Universe.

One of the directions, we have 
followed \cite{nak14,nak15,nak15a1,rog15,rog15a,rog16a,rog16b,rog17} was to analyze 
the effect of {\it dark matter} on the superconducting properties of
materials in order to uncover possible  effects which could be related
to {\it dark sector}. The sharpness of the superconducting transition should 
be helpful to detect even small changes
of e.g., transition temperature due to the presence of the {\it dark matter}. 
Generally it is argued that the {\it dark sector} affects various properties of the systems \cite{pen15,pen17}.    
Studying these changes may contribute to uncover other than gravity effects of {\it dark matter} sector.

As noted earlier one can interpret the second field in action (\ref{sgrav}) 
as {\it the dark sector} coupled to the visible one.
Having in mind that the coupling to the {\it dark sector} changes only the pre-factors of $Q_{(F)}$  
we conclude that in the studied geometry with magnetic field perpendicular to
the plane of graphene it will be very difficult, if possible at all, to detect the
effect of {\it dark matter} experimentally (more details below).  
The situation might change for the geometry with in-plane magnetic field, 
as the recent experimental detection of the mixed gauge-gravitational anomaly suggests \cite{gooth2017}.
This issue is the subject of the  on-going studies.

The observed dependence of transport on $g$ and $\alpha$  can be in principle at least utilized 
in future experiments aiming at the detection of the {\it dark sector}. One possible approach could be the
long-time observations of the properties of well characterized graphene sample. If the  
{\it dark matter} exists, as required by the astrophysical
observations,  so it may be spotted during the annual motion of the Earth~\cite{foo15}-\cite{foo15a} and \cite{freese2013}-\cite{kou14}.
The possible effect of the {\it dark matter} on graphene can 
in principle be detected by the precise and cleverly designed experiments looking at the
annual changes of their transport properties. We rely here on the arguments presented in the aforementioned works, 
where the authors analyze the annual modulations of the {\it dark matter}.
Our additional assumption is that {\it dark matter} is non-homogeneously distributed in the neighborhood 
of the Sun~\cite{frenk2012,jungman1996} and these inhomogeneities can be vital for its detection ~\cite{davoudiasl2012}.
The theoretically expected small value of $\alpha$-coupling constant is an important factor
making the experiments very difficult, but maybe not impossible.




\acknowledgments
MR was partially supported by the grant DEC-2014/15/B/ST2/00089 of the National Science Center  
and KIW by the grant DEC-2014/13/B/ST3/04451.



\end{document}